\documentclass[11pt]{article}
\usepackage{amsmath,amssymb,amsfonts}
\usepackage[dvips]{color}
\usepackage{xcolor}

\usepackage{amsthm}
\usepackage{graphicx}
\usepackage{caption}
\usepackage{subcaption}
\usepackage{bm}
\usepackage{natbib}
\usepackage{rotating}
\usepackage{afterpage}
\usepackage{cancel}
\usepackage{booktabs}
\usepackage{threeparttable}
\usepackage{setspace}
\usepackage{enumitem}
\usepackage{mathabx}
\usepackage{footnote}
\usepackage{mathpazo}

\newcommand{\bc}{\begin{center}}
\newcommand{\ec}{\end{center}}
\newcommand{\be}{\begin{equation}}
\newcommand{\ee}{\end{equation}}
\newcommand{\bea}{\begin{eqnarray}}
\newcommand{\eea}{\end{eqnarray}}
\newcommand{\bean}{\begin{eqnarray*}}
\newcommand{\eean}{\end{eqnarray*}}
\newcommand{\bt}{\begin{tabular}}
\newcommand{\et}{\end{tabular}}

\usepackage{url}
\usepackage{xcolor}
\usepackage[margin=1in]{geometry}
\usepackage{graphicx}

\numberwithin{theorem}{section}

\newcounter{saveeqn}

\usepackage{adjustbox}
\usepackage{hyperref}

\hypersetup{
	colorlinks = true,
	citecolor = [rgb]{0.6824, 0.0431, 0.1647},
	linkcolor = [rgb]{0.6824, 0.0431, 0.1647},
	urlcolor =  [rgb]{0.1137, 0.2745, 0.5020}}

\begin{document}

\title{\bf The Promise of Time-Series Foundation Models for Agricultural Forecasting: Evidence from Commodity Prices}

\author{
	Le Wang\thanks{AAEC, Virginia Tech, Blacksburg, VA, 24061, USA. Email: Le.Wang.Econ@gmail.com}\\ {\small {\em Virginia Tech}} \\
	\and
	Boyuan Zhang\thanks{Amazon, Bellevue, WA, 98004, USA. Email: zhang.boyuan@hotmail.com. This research was conducted independently and is not related to the author's position at Amazon. The views expressed are solely those of the authors and do not reflect the position of Amazon or its subsidiaries. We thank Olga Isengildina Massa for her useful comments.} \\
	{\small{\em Amazon}} 
}

\date{\vspace{-0.5cm} \today}

\maketitle

% \vspace{-10mm}

\begin{abstract}
    Forecasting agricultural markets remains challenging due to nonlinear dynamics, structural breaks, and sparse data. A long-standing belief holds that simple time-series methods outperform more advanced alternatives. This paper provides the first systematic evidence that this belief no longer holds with modern time-series foundation models (TSFMs). Using USDA ERS monthly commodity price data from 1997-2025, we evaluate 17 forecasting approaches across four model classes, including traditional time-series, machine learning, deep learning, and five state-of-the-art TSFMs (Chronos, Chronos-2, TimesFM 2.5, Time-MoE, Moirai-2), and construct annual marketing year price predictions to compare with USDA's futures-based season-average price (SAP) forecasts. We show that zero-shot foundation models consistently outperform traditional time-series methods, machine learning, and deep learning architectures trained from scratch in both monthly and annual forecasting. Furthermore, foundation models remarkably outperform USDA's futures-based forecasts on three of four major commodities despite USDA's information advantage from \emph{forward-looking} futures markets. Time-MoE delivers the largest accuracy gains, achieving 54.9\% improvement on wheat and 18.5\% improvement on corn relative to USDA ERS benchmarks on recent data (2017-2024 excluding COVID). These results point to a potential paradigm shift in agricultural forecasting.
\end{abstract}

\noindent \textbf{JEL Code}: C53, Q11, Q13, C45

\noindent \textbf{Key words}: Time Series Foundation Models, Agricultural Price Forecasting, Marketing Year Average Price, Machine Learning, Deep Learning

% \noindent \textbf{Disclaimer}: This paper and its contents are not related to Amazon and do not reflect the position of the company and its subsidiaries.
%}

\thispagestyle{empty}
\setcounter{page}{0}

\clearpage
%====================================================================
\section{Introduction}
\label{sec:introduction}

% Forecasting agricultural variables presents an enduring and complex challenge. Agricultural markets are shaped by a confluence of volatile and often unpredictable factors -- abrupt weather shocks, evolving political environments, shifting trade policies, and structural changes in global supply chains. These sources of uncertainty can lead to nonlinear dynamics, regime shifts, and structural breaks that undermine the stability of traditional time series relationships. The challenge is further compounded by data limitations, especially in specialized agricultural markets where high-frequency or long-span datasets are scarce, making it difficult to train and validate more sophisticated forecasting models.

% Among all agricultural indicators, prices are among the most challenging to predict because they reflect the aggregate outcome of multiple interacting forces. Prices respond not only to short-term fluctuations in supply and demand, but also to long-run structural factors such as technological change, global trade realignments, and evolving policy regimes. They encapsulate the cumulative effects of production cycles, climate variability, and market sentiment -- each operating at different temporal and spatial scales. As a result, even minor shifts in underlying conditions can generate large and persistent deviations from historical trends.  Yet, accurate forecasts of agricultural prices are crucial for decision-making by farmers, agribusinesses, and policymakers. 

Forecasting agricultural commodity prices presents an enduring and complex challenge. Prices are shaped by a confluence of volatile and often unpredictable factors -- abrupt weather shocks, evolving political environments, shifting trade policies, and structural changes in global supply chains. These sources of uncertainty generate nonlinear dynamics, regime shifts, and structural breaks that undermine the stability of traditional time series relationships. Prices respond not only to short-term supply and demand fluctuations, but also to long-run structural factors such as technological change and market sentiment, each operating at different temporal scales. Even minor shifts in underlying conditions can generate large and persistent deviations from historical trends. The challenge is compounded by data scarcity. Specialized agricultural markets often lack the high-frequency or long-span datasets needed to train and validate sophisticated forecasting models. Yet, accurate price forecasts are crucial for decision-making by farmers, agribusinesses, and policymakers.

Traditional forecasting methods for agricultural prices have employed two main approaches. USDA Economic Research Service (ERS) uses a futures-plus-basis methodology \citep{hoffman2005forecasting,hoffman2007forecasting,hoffman2015, hoffman2018cotton}, combining \emph{forward-looking} market expectations with historical price differentials. Academic and industry forecasters, on the other hand, have employed \emph{backward-looking} time series models such as ARIMA and VAR \citep{box1970time, sims1980macroeconomics}, which leverage statistical patterns in \emph{historical} prices. These approaches, while effective in many contexts, cannot capture complex nonlinear patterns and long-range dependencies in price dynamics due to their linear structure. Recent years have witnessed rapid advances in time series forecasting techniques, moving beyond traditional statistical models toward a new generation of machine learning, deep learning, and more recently, time-series foundational models (TSFM). These approaches promise to capture nonlinear dependencies, high-dimensional feature interactions, and cross-series relationships that conventional econometric models often miss. In particular, time-series foundation models, trained on vast and diverse datasets across domains, claim to offer unprecedented generalization capabilities -- adapting to new contexts with minimal or even without any retraining. However, whether these promises translate effectively to the agricultural sector remains largely unexplored. Agricultural data pose distinctive challenges as discussed above. Consequently, it is not yet clear whether these advanced models can deliver meaningful gains in predictive performance over simpler, well-calibrated econometric models -- especially those USDA's futures-based forecasts with information advantage from forward-looking futures markets -- in such specialized settings. 

Our paper directly addresses this knowledge gap by providing the first comprehensive evaluation of time-series foundation models for agricultural price forecasting. Prior evaluations of agricultural forecasting methods suffer from heterogeneity in methodologies, commodities, and sample periods, making systematic comparison difficult \citep{isengildina2024optimal}. We address this limitation through a unified evaluation framework: 17 models across four commodities using identical data, forecast horizons, and metrics, enabling direct performance comparisons. From a methodological perspective, our study contributes to understanding when and why foundation models succeed in specialized domains. Agricultural prices present an ideal test case: limited training data (200-400 monthly observations), strong domain-specific patterns (biological production cycles, policy-driven structural breaks), and an established operational benchmark (USDA ERS forecasts). Our findings that zero-shot foundation models outperform domain-specialized methods have implications for forecasting in other data-scarce economic domains. 

Our analysis proceeds in two complementary stages. First, we conduct a comprehensive comparison of forecasting performance across a wide spectrum of models, ranging from traditional econometric approaches (such as ARIMA and exponential smoothing) to machine learning (e.g., random forests, gradient boosting), deep learning (e.g., LSTM, DeepAR), and the latest time-series foundation models. This stage focuses on monthly price forecasting up to 12 months, enabling us to assess each model's ability to capture short-term fluctuations, nonlinear dependencies, and evolving market dynamics.

Second, we benchmark the forecasting performance of these models against a policy-relevant metric: the Marketing Year Average (MYA) price, calculated as the season-average of monthly prices weighted by marketing percentages (hence USDA's term "Season-Average Price" or SAP). USDA ERS produces SAP forecasts using a transparent, model-based methodology (futures-plus-basis) that serves as a key input to official WASDE reports. We use ERS SAP forecasts as our benchmark because they are publicly available and replicable, unlike WASDE SAP forecasts which incorporate confidential expert judgment and analyst research. Throughout this paper, we use "MYA" to refer to the price metric itself and our model forecasts of it, while "SAP" refers specifically to USDA ERS forecasts of the same metric. Both terms describe the identical concept: the marketing-year weighted average of monthly prices. 

We focus on MYA prices for two interrelated reasons: substantive policy relevance and methodological significance as a challenging forecasting target. From a substantive perspective, MYA prices are among the most policy-relevant indicators in U.S. agriculture. They represent the weighted average of prices received by farmers over the marketing year. ERS SAP forecasts serve as key inputs to official WASDE SAP forecasts, which in turn are used to calculate Price Loss Coverage (PLC) and Agricultural Risk Coverage (ARC) payment rates for program commodities \citep{hoffman2018cotton}. WASDE reports guide production, investment, and trading decisions by farmers, agribusinesses, policymakers, and other stakeholders. With billions of dollars in farm program payments depending on these price forecasts, even modest improvements in forecast accuracy can have substantial economic impacts. 

From a methodological perspective, MYA prices offer a stringent test case for evaluating forecasting models. The USDA ERS benchmark incorporates futures prices from actively traded markets where participants have strong financial incentives to form accurate expectations \citep{fama1970efficient}. A substantial body of research demonstrates the superiority of futures-based approaches over statistical time-series methods across diverse commodities and horizons \citep{kastens1998, manfredo2004, roache2011, hoffman2015, etienne2019, figuerola2021}. This performance advantage arises because futures markets aggregate dispersed information about anticipated supply, demand, weather, and policy conditions \citep{adjemian2020}, making them difficult to outperform with backward-looking statistical models. The scope for improvement thus appears limited, setting a high bar for alternative forecasting methods. This makes MYA prices an ideal setting for examining whether modern machine learning, deep learning, and time-series foundation models can achieve superior predictive accuracy even when competing against informationally efficient market-based benchmarks.

Our evaluation indeed reveals that the conventional wisdom that ``simple models forecast best'' holds for past deep learning methods -- but breaks down decisively with modern pre-trained time-series foundation models, which consistently outperform both traditional and advanced alternatives. Specifically, using data from the USDA ERS on corn, soybeans, wheat, and cotton prices spanning 1997--2025, we evaluate 17 forecasting methods across four categories: traditional time series (ARIMA, ETS, STL, Prophet, Naive), machine learning (Random Forest, XGBoost), deep learning (LSTM, N-BEATS, TFT, DeepAR), and foundation models (Chronos, Chronos-2, TimesFM 2.5, Time-MoE, Moirai-2). The 28-year sample evaluation period spans multiple agricultural cycles, major policy changes, and pronounced market disruptions -- including the 2008 financial crisis, the 2012 drought, and the COVID-19 pandemic -- and features substantial cross-commodity variation in price volatility. Through a unified evaluation framework with 1,088 forecasts across 64 train-test splits, we find that foundation models achieve superior performance in monthly price forecasting, with all five ranking in the top five positions. Time-MoE leads with the smallest RMSE and MAE, followed by Chronos and Chronos-2. Deep learning models trained from scratch consistently underperform, ranking 10th--16th despite extensive hyperparameter optimization. The Naive model (rank 6) outperforms all deep learning models, demonstrating that simpler approaches generalize better with limited training data.

For MYA forecasting, foundation models demonstrate strong performance across major commodities despite using only historical prices while USDA incorporates forward-looking futures market information. On wheat, 14 of 17 models outperform USDA's SAP prediction, led by Time-MoE with 54.9\% improvement in mean absolute error (MAE). On corn and soybeans, foundation models achieve selective but meaningful advantages, with Time-MoE improving 18.5\% on corn and TimesFM improving 6.9\% on soybeans. Cotton remains challenging, with USDA maintaining superior accuracy. Our analysis reveals that smaller mixture-of-experts architectures outperform larger dense transformers, that zero-shot foundation models are competitive with specialized baselines without requiring domain-specific training, and that simpler models generalize better with limited data. These findings have important implications for farm policy design, risk management, and the broader application of foundation models to economics.

%By focusing on MYA prices, our project achieves two broader objectives. First, it establishes a policy-grounded benchmark for evaluating forecasting methods in agriculture -- bridging the gap between methodological innovation and applied policy relevance. Second, it provides critical insights into the conditions under which modern time-series models outperform, guiding both researchers and policymakers in selecting appropriate forecasting tools for different types of agricultural and economic data.

%This paper addresses three fundamental questions: First, how do state-of-the-art time-series foundation models perform compared to traditional methods in forecasting MYA prices in zero-shot mode? Second, are there systematic patterns in model performance across different commodity groups (row crops versus fiber crops) that can inform model selection? Third, what architectural features (model size, mixture-of-experts versus dense transformers) drive foundation model performance on agricultural price forecasting?

%====================================================================
\subsection*{Connections to the Existing Literature}

Our work contributes to three strands of literature: agricultural price forecasting, time series foundation models, and the application of machine learning to economic forecasting.\footnote{The forecasting literature encompasses far more methods than we can evaluate here, including additional statistical approaches (e.g., TBATS, state-space models), machine learning variants (e.g., LightGBM, CatBoost, support vector regression), deep learning architectures (e.g., WaveNet, Informer, Autoformer, PatchTST), and emerging foundation models (e.g., TimeGPT, Lag-Llama, MOMENT). For comprehensive surveys, see \citet{hyndman2018forecasting} for traditional and machine learning approaches, \citet{makridakis2018m4} for comparative evaluation across method classes, \citet{zhang2024llm} for large language models applied to time series, and \citet{kottapalli2025foundation} for recent developments in time series foundation models. Our goal is not exhaustive comparison but rather to demonstrate that foundation models can be useful for agricultural forecasting and that well-established methods across model classes can be competitive with USDA's ERS SAP forecasts.}

\textit{Agricultural price forecasting.} The literature on agricultural commodity price forecasting provides the empirical foundation for our study. Early theoretical work by \citet{Working1949theory} developed the theory of storage linking spot and futures prices, while \citet{fama1970efficient} established that futures prices should be unbiased predictors of future spot prices under market efficiency. Building on these foundations, \citet{tomek1997commodity} provided empirical evidence that futures prices serve as efficient predictors of cash prices for agricultural commodities, establishing the theoretical basis for the USDA ERS futures-plus-basis methodology that remains the operational standard we benchmark against.

Subsequent research has examined the accuracy of government forecasts. \citet{sanders2008forecasting} documented significant market reactions to USDA reports for wheat, soybeans, and hogs, while \citet{isengildina2011comparison} compared USDA forecasts with private sector predictions across corn, soybeans, wheat, and cotton, finding that USDA maintains competitive accuracy despite using simpler methodologies. Systematic reviews \citep{isengildina2024optimal} document that USDA forecasts consistently outperform naive and time-series alternatives and achieve accuracy comparable to market-based forecasts, establishing them as a demanding benchmark for methodological innovation. This result motivates our focus on whether modern methods can improve upon this established benchmark.

The USDA ERS futures-plus-basis methodology has evolved through systematic research by Hoffman and colleagues. \citet{hoffman2005forecasting} developed the initial futures-based model for corn, demonstrating that adding 5-year average basis to futures prices produces reliable forecasts. \citet{hoffman2007forecasting} extended this approach to soybeans and wheat, finding that 5-year moving averages effectively address structural changes while maintaining forecast precision. \citet{hoffman2018cotton} adapted the methodology for cotton using 7-year Olympic averages (excluding maximum and minimum values) to improve basis forecast stability. This methodological development reflects broader empirical evidence that futures-based forecasts consistently outperform time-series alternatives across agricultural commodities \citep{kastens1998, manfredo2004, roache2011, hoffman2015, etienne2019, figuerola2021}. The performance advantage arises because futures markets aggregate dispersed information about anticipated supply, demand, weather, and policy conditions \citep{adjemian2020}, making them difficult to outperform with backward-looking statistical models.

Machine learning applications to agricultural prices have yielded mixed results. \citet{zhang2003time} showed that hybrid ARIMA-neural network models can capture both linear and nonlinear price patterns. More recently, \citet{zelingher2024agricaf} proposed AGRICAF, an explainable framework incorporating exogenous drivers (energy prices, fertilizer costs, stock levels) to forecast maize, soybean, and wheat prices up to one year ahead. Their covariate-rich approach contrasts with ours: we evaluate whether foundation models can achieve competitive accuracy using only historical prices in a purely univariate setting, mimicking practitioners without access to specialized databases. Despite these advances, agricultural forecasting remains challenging due to extreme price volatility \citep{wright2011economics}, structural breaks from policy changes \citep{goodwin2000forecasting}, and data scarcity -- monthly series span only 20--30 years, providing at most a few hundred observations. This data limitation has important implications for model selection, as discussed below.

\textit{Traditional time series and machine learning methods.} Beyond domain-specific applications, a rich methodological literature has developed general-purpose forecasting techniques that we evaluate in this study. Classical statistical methods remain competitive in data-scarce settings: \citet{holt1957forecasting} and \citet{winters1960forecasting} developed exponential smoothing for trend and seasonality, \citet{cleveland1990stl} introduced STL decomposition using locally weighted regression, and \citet{taylor2018forecasting} proposed Prophet for automated forecasting with changepoint detection. Tree-based machine learning methods have also become standard tools: \citet{breiman2001random} introduced random forests for capturing nonlinear relationships, while \citet{chen2016xgboost} developed gradient boosting methods that achieve strong performance through sequential error correction.

\textit{Deep learning for time series forecasting.} While tree-based methods can work with limited data, deep learning architectures face more severe constraints. \citet{goodfellow2016deep} established that deep learning models require thousands of training samples to avoid overfitting -- a threshold rarely met in economic applications and certainly not in commodity price forecasting where monthly series contain only 200--400 observations. \citet{hochreiter1997long} introduced Long Short-Term Memory (LSTM) networks that can capture long-term dependencies through gated recurrent units. Several architectures have been developed specifically for time series: \citet{oreshkin2020nbeats} developed N-BEATS using stacks of residual blocks to decompose forecasts into interpretable trend and seasonal components, achieving strong performance on M4 competition; \citet{lim2021temporal} proposed Temporal Fusion Transformers (TFT) combining LSTM encoders with multi-head self-attention mechanisms for learning long-range dependencies; and \citet{salinas2020deepar} introduced DeepAR for probabilistic forecasting using autoregressive recurrent networks with distributional outputs. Despite these architectural innovations, \citet{makridakis2018statistical} found that simpler statistical methods often outperform complex deep learning models when data are limited, particularly in economic applications.

\textit{Time series foundation models.} The data requirements of deep learning have motivated a new paradigm: foundation models that leverage pre-training on massive external datasets to overcome sample size limitations in target domains. Foundation models represent a paradigm shift in machine learning, where large models pre-trained on diverse datasets can be applied to new tasks with minimal adaptation \citep{bommasani2021opportunities}. In time series forecasting, recent foundation models have achieved strong performance across diverse benchmarks. We evaluate five representative models in this study. \citet{ansari2024chronos} introduced Chronos, which adapts language modeling techniques to time series by tokenizing numerical values and applying transformer architectures with 200 million parameters. \citet{ansari2025chronos2} extended this work with Chronos-2, incorporating mixture-of-experts architecture with 120 million parameters. \citet{das2024timesfm} developed TimesFM using decoder-only transformers with frequency-aware encodings, pre-trained on over 100 billion time points. \citet{jin2024timemoe} proposed Time-MoE, a mixture-of-experts architecture with only 50 million parameters through sparse activation patterns. \citet{aksu2025moirai2} introduced Moirai-2, using quantile forecasting and multi-token prediction for improved efficiency.

These foundation models have demonstrated impressive zero-shot performance on standard benchmarks like M4 and M5 competitions. However, recent work has questioned whether TSFMs truly exhibit ``foundational'' properties. \citet{karaouli2025tsfm} argued that zero-shot capabilities are significantly tied to pre-training domains, and that fine-tuned foundation models do not consistently outperform smaller dedicated models relative to their increased parameter count. This skepticism is particularly relevant for specialized domains like agricultural economics, which present unique challenges: strong seasonality from biological production cycles, structural breaks from policy changes, and high volatility from weather and trade shocks. Whether foundation models can generalize to these domain-specific patterns without fine-tuning is an open empirical question that our study addresses. Our findings contribute to this debate by documenting that in agricultural price forecasting, zero-shot TSFMs substantially outperform both traditional methods and deep learning models trained from scratch, with one of the most parameter-efficient foundation models (Time-MoE) achieving the best overall performance.

\vspace{.1in}

The remainder of the paper proceeds as follows. Section~\ref{sec:data} presents the data and USDA ERS forecast methodology. Section~\ref{sec:methods} describes the forecasting methods and model taxonomy. Section~\ref{sec:results} reports empirical results for all models. Section~\ref{sec:discussion} discusses practical implications and limitations. Section~\ref{sec:conclusion} concludes.

%====================================================================
\section{Data}
\label{sec:data}

%====================================================================
\subsection{Data Sources and Coverage}

Our analysis uses monthly price data from the USDA Economic Research Service (ERS) database for SAP Forecasts, covering major U.S. agricultural commodities that collectively represent over 200 million planted acres annually. Table \ref{tab:data_summary} provides an overview of the data structure, variable definitions, and sources.

\begin{table}[ht!]
   \centering
   \caption{Data Summary: Variables, Definitions, and Sources}
   \label{tab:data_summary}
   \footnotesize
   \adjustbox{max width=1.0\textwidth}{ 
      \begin{tabular}{lp{4cm}lll}
         \toprule
         \textbf{Variable} & \textbf{Definition} & \textbf{Unit} & \textbf{Frequency} & \textbf{Source} \\
         \midrule
         \multicolumn{5}{l}{\textit{Panel A: Primary Variables}} \\
         \midrule
         Monthly Price & Price received by farmers & \$/bu (corn, soy, wheat) & Monthly & USDA NASS \\
         & for commodity sales & cents/lb (cotton) & & Agricultural Prices \\
         \midrule
         \multicolumn{5}{l}{\textit{Panel B: Benchmark Forecasts}} \\
         \midrule
         USDA SAP Forecast & ERS SAP forecast & \$/bu or cents/lb & Monthly updates & USDA ERS \\
         & (futures + basis method) & & (annual target) & Season-Avg Forecasts \\
         \midrule
         \multicolumn{5}{l}{\textit{Panel C: Auxiliary Variables}} \\
         \midrule
         Futures Price & Nearby futures contract & \$/bu or cents/lb & Daily & LSEG Data \& Analytics \\
         & settlement price & & & (formerly Refinitiv) \\
         Basis & Cash price - futures price & \$/bu or cents/lb & Monthly & USDA ERS \\
         & (5-year or 7-year average) & & & (calculated) \\
         Marketing \% & Proportion of annual production  marketed in each month & Percentage & Monthly & USDA NASS Agricultural Prices \\
         \bottomrule
      \end{tabular}
   }
   \vskip 2mm
   \begin{minipage}{0.98\textwidth}
      {\footnotesize \textit{Notes}: This table summarizes all variables used in the analysis. Panel A shows the primary variable (monthly prices received by farmers) used to generate forecast. Panel B shows the benchmark forecast used for comparison. ERS SAP forecasts are generated weekly using data available each Thursday; files are updated monthly and posted 1 business day after WASDE report release. Panel C shows auxiliary variables used only by baseline methods (USDA futures-plus-basis approach). Marketing percentages are fixed weights (5-year average for corn/soybeans/wheat, 7-year Olympic average for cotton) used to aggregate monthly prices to MYA. All data are publicly available from USDA sources. See Appendix~\ref{sec:app_data} for complete data access details and download links.\par}
   \end{minipage}
\end{table}

We use the latest dataset as of 12/10/2025, which includes four major U.S. agricultural commodities: corn (September 1997 -- August 2025), soybeans (September 1997 -- August 2025), wheat (June 1997 -- August 2025), and cotton (August 1997 -- August 2025). Each commodity follows its specific marketing year calendar\footnote{Marketing years are 12-month periods aligned with each commodity's harvest and marketing cycle, used by USDA for reporting production, sales, and price statistics.} defined by USDA: corn and soybeans (September--August), wheat (June--May), and cotton (August--July). Throughout this paper, we denote marketing years by their starting year. For example, MY 2023 for corn refers to the period September 2023 through August 2024. These marketing years align with harvest timing and traditional marketing patterns for each crop.

\begin{figure}[ht!]
   \centering
   \caption{Monthly Commodity Prices (1997-2025)}
   \label{fig:timeseries}
   \includegraphics[width=0.95\textwidth]{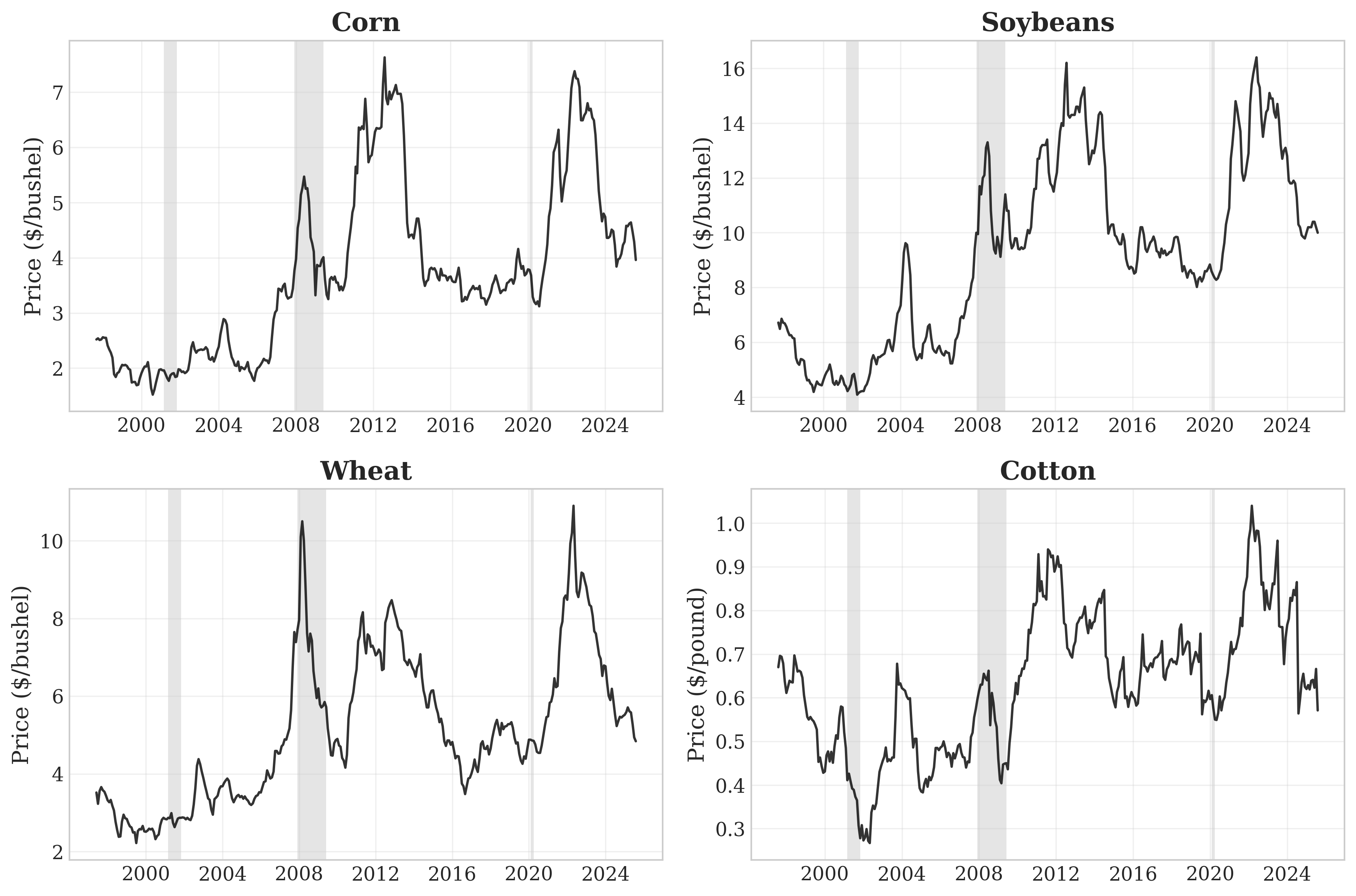}
   \begin{tablenotes}
   \small
   \item \textit{Notes}: Time series of monthly commodity prices from USDA ERS. Gray shaded areas indicate NBER recession periods (2001 dot-com recession, 2007-2009 Great Recession, 2020 COVID-19 recession). All commodities exhibit substantial volatility and structural breaks that challenge traditional forecasting methods.
   \end{tablenotes}
\end{figure}

Figure \ref{fig:timeseries} displays the complete monthly prices for all four commodities, illustrating the diverse price dynamics and volatility patterns that foundation models must capture. The 28-year period encompasses multiple agricultural cycles, major policy changes,\footnote{Major Farm Bill legislation during our sample period includes the 2002 Farm Security and Rural Investment Act, 2008 Food, Conservation, and Energy Act, 2014 Agricultural Act, and 2018 Agriculture Improvement Act \citep{zulauf2014farm, schnitkey2019plc}. These policy changes affected reference prices, payment structures, and program eligibility, creating structural breaks in price dynamics.} and significant market disruptions including the 2008 financial crisis, 2012 drought, and COVID-19 pandemic. Price volatility varies substantially across commodities: soybeans exhibit the highest coefficient of variation (standard deviation divided by mean, 44\%), followed by corn (38\%), wheat (32\%), and cotton (25\%).
 
%====================================================================
\subsection{USDA ERS Season-Average Price Forecasts}
\label{sec:mya_prices}

The USDA ERS produces futures-based season-average price (SAP) forecasts that serve as key inputs to the official WASDE consensus forecasts \citep{poghosyan2025cotton}. WASDE reports combine analyst expertise, econometric models, and market intelligence to generate official price projections. To prevent market manipulation, WASDE preparation follows strict confidentiality protocols, with no information released until the scheduled publication time. In contrast, the ERS SAP forecasts are publicly documented and methodologically transparent, making them suitable as a replicable benchmark. The USDA ERS has produced SAP forecasts since 2003 using a futures-plus-basis methodology in two steps \citep{hoffman2005forecasting, hoffman2007forecasting,hoffman2015, hoffman2018cotton}.

The ERS methodology first generates monthly price forecasts using futures prices and historical basis patterns. The forecast for month $k$ in year $t$ is generated as:
\begin{align}
   \widehat{P}_{k,t} &= F_{h(k),t,i} + \bar{B}_{k,t}, \label{eq:hoffman_model}
\end{align}
where $\widehat{P}_{k,t}$ is the predicted farm price in calendar month $k$ and year $t$, $F_{h(k),t,i}$ is the futures price for the nearby contract expiring in month $h(k)$ observed at forecast horizon $i$, and $\bar{B}_{k,t}$ is the historical average basis.\footnote{For corn, the nearby contract mapping follows the harvest cycle: December futures for October-November prices, March futures for December-February, May futures for March-April, July futures for May-June, and September futures for July-August. Similar mappings apply to other commodities based on their respective marketing years and contract availability.} This methodology forecasts each month directly using the corresponding nearby futures contract plus historical basis, avoiding error accumulation across the 12-month horizon. The basis reflects local supply-demand conditions, transportation costs, and storage premiums that cause cash prices to deviate from futures prices. The basis averages use 5-year moving windows for corn, soybeans, and wheat, and 7-year Olympic averages for cotton to address structural changes while maintaining forecast reliability.\footnote{This creates information asymmetry in our comparison: USDA forecasts incorporate forward-looking futures prices, while our models use only historical cash prices. We discuss the implications of this design choice in Section~\ref{sec:discussion}. Briefly, this represents a more stringent test for our models -- outperforming futures-based forecasts using only backward-looking information demonstrates genuine forecasting ability rather than simply incorporating the same market signals.}

Monthly forecasts are then aggregated to the Marketing Year Average (MYA), which represents the weighted average price over the 12-month marketing year:
\begin{align} \label{eq:mya_price}
   \widehat{\text{MYA}}_t &= \sum_{k=1}^{12} \widehat{P}_{k,t} \times w_k,
\end{align}
where $\widehat{P}_{k,t}$ is the monthly price predicted by Equation (\ref{eq:hoffman_model}) and $w_k$ is the monthly marketing percentage -- the proportion of annual production marketed in month $k$, with $\sum_{k=1}^{12} w_k = 1$. The actual MYA is calculated using actual farmer-received prices and actual marketing percentages within that marketing year. Marketing percentages reflect the typical temporal pattern of commodity sales throughout the year. For example, corn marketing percentages are highest immediately post-harvest (September--November: 45\% of annual sales) and decline through the marketing year as on-farm storage is depleted. For forecasting purposes, each month's marketing percentage uses the 5-year average of that same month's historical percentage for corn, soybeans, and wheat, and a 7-year Olympic average (dropping the highest and lowest values) for cotton.

%====================================================================
\section{Forecasting Methods}
\label{sec:methods}

%====================================================================
\subsection{Model Taxonomy}

We evaluate 17 forecasting models across four methodological paradigms: traditional time-series approaches (6 models), machine learning (2 models), deep learning (4 models), and foundation models (5 models). This section outlines the basic structure and intuition behind each method, along with brief notes on implementation and hyperparameter selection where relevant. After introducing each method, we also briefly discuss its advantages and disadvantages. Our goal is not to provide an exhaustive technical treatment but rather to offer concise descriptions and direct readers to key references for deeper study. Full implementation details are provided in the Appendix \ref{sec:app_tsfm} and \ref{sec:app_cv}, and more advanced discussions of each method can be found in the papers cited below.\footnote{All models are trained and evaluated separately for each commodity using purely univariate forecasting -- each model uses only the target commodity's own price history. This design follows standard practice in applied forecasting \citep{makridakis2018m4} and provides a fair comparison of model architectures under comparable temporal constraints. Cross-commodity pooling and covariate-augmented forecasting represent natural extensions for future work.} In what follows, $\hat{P}_{t+h}$ denotes the $h$-step \emph{forecast} of the price at time $t+h$, and $P_t$ is the \emph{observed} price at time $t$.

%====================================================================
\subsubsection{Traditional Time Series Models}

Traditional time series methods model temporal dependence through explicit statistical assumptions about trend, seasonality, and autocorrelation. We implement six models representing different approaches to decomposing and forecasting price dynamics. 

We start with the simplest models. The \emph{Naive} model assumes prices follow a random walk, forecasting $\hat{P}_{t+h} = P_t$ for all horizons $h$. The \emph{Seasonal Naive model} exploits annual patterns by forecasting $\hat{P}_{t+h} = P_{t+h-12}$, using the price from the same month in the previous year. These simple benchmarks serve as important baselines in forecasting competitions and applied work \citep{makridakis2018m4}, where they sometimes surprisingly outperform more complex models. Both models are computationally trivial with no parameters to tune, providing robust baselines for comparison. They cannot capture complex patterns or structural breaks, and Seasonal Naive requires at least one full year of historical data.

\emph{SARIMA} (Seasonal AutoRegressive Integrated Moving Average) extends the foundational ARIMA framework developed by \citet{box1970time} to handle seasonal patterns. The original ARIMA model revolutionized time series analysis by providing a unified framework for modeling non-stationary data through differencing, while the seasonal extension enables explicit modeling of recurring annual patterns common in agricultural prices. SARIMA remains the workhorse of applied forecasting due to its interpretable parameters and well-understood statistical properties. The model decomposes prices into autoregressive (AR), differencing (I), and moving average (MA) components, with seasonal counterparts. The general SARIMA$(p,d,q)(P,D,Q)_s$ model is compactly written as follows:
\begin{align}
   \Phi(B^s)\phi(B)(1-B)^d(1-B^s)^D P_t &= \Theta(B^s)\theta(B)\epsilon_t,
\end{align}
where $B$ is the backshift operator ($BP_t = P_{t-1}$), $s=12$ is the seasonal period, $\phi(B)$ and $\Phi(B^s)$ are autoregressive polynomials for non-seasonal and seasonal components, $\theta(B)$ and $\Theta(B^s)$ are moving average polynomials, and the differencing operators $(1-B)^d$ and $(1-B^s)^D$ remove non-seasonal and seasonal trends respectively. For $h$-step ahead forecasting, SARIMA generates predictions recursively: the model first forecasts $\hat{P}_{t+1}$, then uses this forecast as input to predict $\hat{P}_{t+2}$, continuing until reaching horizon $h$. The algorithm selects optimal orders based on information criterion separately for each commodity-split combination, balancing model fit against complexity. SARIMA benefits from a well-established statistical foundation with interpretable parameters and the ability to handle non-stationarity through differencing, but it assumes linear relationships, is sensitive to outliers, and requires sufficient historical data for reliable parameter estimation.
% We use automatic AIC-based order selection, testing $p,q,P,Q \in \{0,1,2\}$ and $d,D \in \{0,1\}$. 

\emph{Exponential Smoothing} (ETS) traces its origins to \citet{holt1957forecasting} and \citet{winters1960forecasting}, who developed methods for forecasting data with trend and seasonality. The Holt-Winters method remains widely used because it explicitly models trend and seasonality as separate components within a unified framework, allowing each to evolve independently while contributing to the forecast. This structural decomposition is intuitive for agricultural prices, where long-term trends (from inflation or productivity changes) and seasonal patterns (from planting and harvest cycles) operate through distinct mechanisms. The method recursively updates level, trend, and seasonal components with exponential weights on past observations. The additive seasonal model updates are:
\begin{align}
   \ell_t &= \alpha(P_t - s_{t-s}) + (1-\alpha)(\ell_{t-1} + b_{t-1}), \quad \text{(level)} \\
   b_t &= \beta(\ell_t - \ell_{t-1}) + (1-\beta)b_{t-1}, \quad \text{(trend)} \\
   s_t &= \gamma(P_t - \ell_t) + (1-\gamma)s_{t-s}, \quad \text{(seasonal)}
\end{align}
where $P_t$ is the observed price at time $t$, $\ell_t$ represents the level (baseline price), $b_t$ is the trend slope (rate of price change), $s_t$ is the seasonal component (recurring monthly patterns), $s_{t-s}$ is the seasonal component from the same season in the previous year, $s=12$ is the seasonal period, and $\alpha,\beta,\gamma \in [0,1]$ are smoothing parameters that control how much weight is given to recent observations versus historical patterns. The $h$-step ahead forecast is:
\begin{align}
   \hat{P}_{t+h} &= \ell_t + hb_t + s_{t-s+((h-1) \bmod s)+1},
\end{align}
where the seasonal index cycles through the most recent seasonal estimates. The ETS model also has a multiplicative version, which follows similar logic; we do not present it here. We grid search over trend type (additive, multiplicative, none) and seasonal type (additive, multiplicative, none), selecting the best specification via validation RMSE. Exponential Smoothing is adaptive to recent changes, computationally efficient, and handles both additive and multiplicative seasonality. The method uses an exponential weighting scheme, can overreact to recent shocks, and requires careful specification of trend and seasonal types.

\emph{STL} (Seasonal-Trend decomposition using Loess), developed by \citet{cleveland1990stl}, separates time series into interpretable components using robust local regression. Like Exponential Smoothing, STL explicitly models trend and seasonality as separate components, but with greater flexibility: the seasonal component can vary over time, and the robust fitting procedure resists distortion from outliers -- a common occurrence in agricultural prices during supply shocks or policy changes. This decomposition-based approach is particularly valuable for understanding the drivers of price movements, as analysts can examine trend and seasonal components separately. STL separates prices into three additive components via iterative smoothing:
\begin{align}
   P_t &= T_t + S_t + R_t,
\end{align}
where $T_t$ is the trend, $S_t$ is the seasonal component with $\sum_{i=1}^s S_i = 0$, and $R_t$ is the residual. The decomposition uses locally weighted regression (Loess) with window size parameters $n_s$ (seasonal) and $n_t$ (trend). For forecasting, each component is extrapolated separately. The seasonal component repeats the most recent cycle:
\begin{align}
   \hat{S}_{t+h} &= S_{t-s+((h-1) \bmod s)+1}.
\end{align}
The trend component is forecasted using Holt's linear method applied to the extracted trend series $\{T_1, \ldots, T_t\}$:
\begin{align}
   \hat{T}_{t+h} &= \ell_t^T + h \cdot b_t^T,
\end{align}
where $\ell_t^T$ and $b_t^T$ are the level and slope estimated from the trend component. The final forecast combines these:
\begin{align}
   \hat{P}_{t+h} &= \hat{T}_{t+h} + \hat{S}_{t+h}.
\end{align}
We use robust STL with grid search over $n_s \in \{7, 13, 25, 35\}$ and $n_t \in \{\text{None}, 13, 25, 51\}$ and restrict to additive decomposition.

\emph{Prophet}, developed by \citet{taylor2018forecasting} at Facebook (now Meta), was designed for business forecasting at scale with minimal manual intervention. The model's key innovation is automatic changepoint detection, which identifies structural breaks in the trend without requiring analysts to specify them in advance. For agricultural prices, this capability is valuable because policy changes, trade disruptions, or technological shifts can cause abrupt trend changes that traditional models struggle to capture. Prophet also provides uncertainty intervals through Bayesian inference, enabling probabilistic forecasting. The model implements an additive decomposition with three main components:
\begin{align}
   P_t &= g_t + s_t + h_t + \epsilon_t.
\end{align}
The trend $g_t$ is a piecewise linear function allowing for multiple changepoints:
\begin{align}
   g_t &= (k + \mathbf{a}_t^T \boldsymbol{\delta})t + (m + \mathbf{a}_t^T \boldsymbol{\gamma}),
\end{align}
where $k$ is the base growth rate, $m$ is the offset (the constant in the intercept of the trend function), $\mathbf{a}_t \in \{0,1\}^J$ indicates which of the $J$ changepoints have occurred by time $t$, $\boldsymbol{\delta}$ are slope adjustments at changepoints, and $\boldsymbol{\gamma}$ are corresponding intercept adjustments that ensure continuity. The seasonal component uses Fourier series:
\begin{align}
   s_t &= \sum_{n=1}^N \left[a_n \cos\left(\frac{2\pi nt}{P}\right) + b_n \sin\left(\frac{2\pi nt}{P}\right)\right],
\end{align}
where $P=12$ is the period and $N=10$ is the number of Fourier terms. 

Bayesian inference via Stan estimates parameters with priors $\delta_j \sim \text{Laplace}(0, \tau)$ for changepoint regularization and $a_n, b_n \sim \text{Normal}(0, \sigma_s)$ for seasonality regularization. We grid search over changepoint prior scale $\tau$ (controlling trend flexibility) and seasonality prior scale $\sigma_s$ (controlling the amplitude of seasonal patterns -- larger values allow stronger seasonality). For $h$-step ahead forecasting, Prophet directly evaluates the fitted model at future time points: $\hat{P}_{t+h} = g_{t+h} + s_{t+h} + h_{t+h}$. Since changepoints are only placed within the training period, $\mathbf{a}_{t+h} = \mathbf{a}_t$ for all forecast horizons, meaning the trend extrapolates forward with the final learned slope $k + \mathbf{1}^T\boldsymbol{\delta}$. Prophet automatically detects trend changepoints, handles holidays and special events, and provides uncertainty intervals through its Bayesian framework. The model is computationally expensive due to Stan MCMC sampling, has many hyperparameters to tune, and can overfit with aggressive changepoint detection.

%====================================================================
\subsubsection{Machine Learning Models}

Machine learning methods treat forecasting as supervised learning, using lagged prices and engineered features to predict future values. We implement two \emph{ensemble} tree-based methods -- \emph{Random Forest} and \emph{XGBoost} --  that differ fundamentally in how they construct ensembles: \emph{Random Forest} uses bagging (bootstrap aggregating) where trees are trained independently in parallel, while \emph{XGBoost} uses boosting where trees are trained sequentially with each tree correcting errors from the previous ensemble. While many machine learning methods exist for time series forecasting, we focus on these two methods due to their well-documented forecast accuracy in applied forecasting competitions and economic applications \citep{makridakis2018m4}.

\emph{Random Forest} constructs an ensemble of decision trees \citep{breiman1984cart}, each trained on a bootstrap sample with random feature selection. For time series forecasting, we create feature vectors:
\begin{align}
   \mathbf{x}_t &= [P_{t-1}, P_{t-2}, \ldots, P_{t-L}, \sin(2\pi m/12), \cos(2\pi m/12)],
\end{align}
where $L$ is the number of lags and $m \in \{1,\ldots,12\}$ is the month. The cyclical encoding captures seasonality without imposing ordinality. This single Fourier pair captures the fundamental annual cycle; more complex seasonal patterns could be modeled with additional harmonics, but the lagged features already carry implicit seasonal information when $L \geq 12$.

Each tree $T_k$ is grown by:
\begin{enumerate}
   \item Bootstrap sampling: Draw $n$ samples with replacement from training data
   \item At each node, randomly select $\sqrt{d}$ candidate features from the $d$ total features
   \item Split on the feature and threshold that minimizes the weighted mean squared error across child nodes
   \item Continue splitting until reaching stopping criteria (maximum depth or minimum samples per leaf)
\end{enumerate}
The final prediction averages across $K$ trees: $\hat{P}_{t+1} = \frac{1}{K}\sum_{k=1}^K T_k(\mathbf{x}_t)$. Multi-step forecasts use recursive prediction where $\hat{P}_{t+h}$ becomes a feature for $\hat{P}_{t+h+1}$. We grid search over $L \in \{6,12,18\}$ and $K \in \{100,200\}$. Random Forest handles non-linear relationships, is robust to outliers, requires minimal hyperparameter tuning, and provides feature importance measures. However, since the model lacks explicit trend and seasonality modeling, it requires careful feature engineering for time series applications. For comprehensive treatments of Random Forest, see \citet{breiman2001random} and \citet{hastie2009elements}.

\emph{XGBoost} (Extreme Gradient Boosting), developed by \citet{chen2016xgboost}, implements gradient boosting where each new tree corrects errors from the previous ensemble. The model builds an additive ensemble:
\begin{align}
   \hat{P}_t^{(K)} &= \sum_{k=1}^K \eta f_k(\mathbf{x}_t),
\end{align}
where each $f_k$ is a weak learner -- a shallow regression tree (typically has depth of 3-6) that individually has limited predictive power. The learning rate $\eta \in (0,1]$ shrinks each tree's contribution, and $\mathbf{x}_t$ uses the same lagged features as Random Forest. At each iteration $k$, a new tree $f_k$ is fitted to the negative gradient of the loss function evaluated at the current ensemble prediction, effectively targeting the residual errors. This sequential error correction allows XGBoost to combine many weak learners into a strong predictor. The key innovation is the regularized objective that penalizes model complexity to prevent overfitting. We grid search over $L \in \{6,12,18\}$, $K \in \{100,200\}$, and $\eta \in \{0.05,0.1\}$. Like Random Forest, multi-step forecasts use recursive prediction where $\hat{P}_{t+h}$ becomes a feature for $\hat{P}_{t+h+1}$. For complete technical details on the regularization framework and optimization algorithm, see \citet{chen2016xgboost}.

The key difference between Random Forest and XGBoost lies in their ensemble strategies: Random Forest uses bagging (bootstrap aggregating), training trees independently on bootstrap samples and averaging their predictions to reduce variance; XGBoost uses adaptive boosting, training trees sequentially where each tree focuses on correcting the residual errors of the previous ensemble to reduce bias.

%====================================================================
\subsubsection{Deep Learning Models}

Deep learning models use neural network architectures to learn hierarchical representations from raw time series data. We implement four state-of-the-art architectures, training them from scratch on agricultural price data using the Adam optimizer.

\emph{LSTM} (Long Short-Term Memory) networks, introduced by \citet{hochreiter1997long}, differ fundamentally from standard feedforward neural networks in how they process data. Standard neural networks treat each input independently, making them unsuitable for time series where the order of observations matters. Earlier recurrent neural networks process data one step at a time but tend to ``forget'' older information as new observations arrive, the vanishing gradient problem. LSTMs overcome this limitation by introducing a specialized internal structure that separates long-term memory, captured by a cell state, from short-term memory, summarized in a hidden state. This method has an innovative gating mechanism that decides, at each point in time, what to remember, what to update, and what to forget. This decision is made through three gates that act like switches: one removes outdated information (forget gate), another adds new relevant signals (input gate), and a third determines what information is carried forward to influence future predictions (output gate). By managing information in this way, LSTMs can recognize long-run patterns -- such as recurring seasonal movements in agricultural prices that repeat from year to year, and short-run shocks arising from weather events and temporary market disruptions.

% The key innovation of the LSTM is the gating mechanism that solves the vanishing gradient problem plaguing earlier recurrent networks. LSTMs use three gates -- forget, input, and output -- to selectively retain or discard information at each time step. The forget gate decides what information from the previous cell state to discard; the input gate determines what new information to store; and the output gate controls what information flows to the next time step. This selective memory enables LSTMs to capture dependencies spanning hundreds of time steps, such as year-over-year seasonal patterns in agricultural prices. For complete technical details on the gating mechanism, see \citet{hochreiter1997long}.

We use a single-layer LSTM with grid search over sequence length, hidden size, and training epochs. The final hidden state feeds into a linear layer to produce price forecasts. For multi-step forecasting, we use autoregressive generation: the model predicts $\hat{P}_{t+1}$, then feeds this prediction back as input to generate $\hat{P}_{t+2}$, continuing recursively until reaching the desired horizon $h$. The architecture requires large training datasets (50K+ parameters), is prone to overfitting with small samples, and produces difficult-to-interpret black-box predictions.

\emph{DeepAR} by \citet{salinas2020deepar} is an autoregressive recurrent network designed for probabilistic forecasting. The model uses an LSTM to encode historical prices, then generates forecasts autoregressively: at each future time step, the network outputs parameters of a probability distribution (mean and variance for a Gaussian), samples a value from this distribution, and feeds it back as input to predict the next step. This process repeats until reaching horizon $h$, with multiple sample paths generating prediction intervals. We implement DeepAR with grid search over LSTM architecture parameters. For consistency with other models, we use mean absolute error loss rather than the original distributional loss, which reduces the probabilistic benefits but enables fair comparison. The model requires large training data, and its distributional assumptions may not hold for agricultural prices. For complete details, see \citet{salinas2020deepar}.

Unlike LSTM which learns implicit representations, \emph{N-BEATS} (Neural Basis Expansion Analysis for Time Series) by \citet{oreshkin2020nbeats} is designed for interpretability by explicitly decomposing forecasts into trend and seasonal components. The model outputs the entire forecast horizon $\hat{P}_{t+1:t+h}$ directly in a single forward pass, rather than generating forecasts autoregressively one step at a time. This direct multi-horizon approach avoids error accumulation that plagues recursive methods. The architecture achieves interpretability by expressing forecasts as weighted combinations of basis functions. For trend, the forecast is a polynomial expansion:
\begin{align}
   \hat{y}^{\text{trend}} &= \sum_{i=0}^{p} \theta^{\text{trend}}_i \cdot t^i,
\end{align}
where $t = [0, 1, \ldots, h-1]$ is the forecast horizon vector and $p$ is the polynomial degree; coefficients $\theta^{\text{trend}}_i$ control the level, slope, and curvature of the projected trend. For seasonality, the forecast uses a Fourier series:
\begin{align}
   \hat{y}^{\text{seas}} &= \sum_{i=0}^{\lfloor h/2 \rfloor - 1} \left[ \theta^{\text{seas}}_i \cos\left(\frac{2\pi i t}{h}\right) + \theta^{\text{seas}}_{i + \lfloor h/2 \rfloor} \sin\left(\frac{2\pi i t}{h}\right) \right],
\end{align}
where $\theta^{\text{seas}}_i$ denotes the vector of coefficients that weight the sine and cosine basis functions in the Fourier expansion, determining the amplitude and phase of each seasonal frequency over the forecast horizon. The model works as follows: a multi-layer neural network takes the lookback window of historical prices as input and outputs the expansion coefficients $\theta$ ($\theta^{\text{trend}}, \theta^{\text{seas}}$). These learned coefficients are then multiplied by the fixed basis functions (polynomials or Fourier terms) to produce the forecast. The neural network's role is to learn which combination of basis functions best captures the patterns in the data -- the basis functions provide interpretable structure while the network provides the flexibility to adapt to different time series.

We implement N-BEATS with a simplified architecture adapted for agricultural data: fewer processing layers and smaller hidden dimensions than the default specification to prevent overfitting. N-BEATS provides interpretable decomposition into trend and seasonality components and achieved strong performance on the M4 competition \citep{makridakis2018m4}. However, the model requires substantial training data (200K+ parameters), needs architecture simplification for small samples, and hence is computationally intensive. For complete architectural details, see \citet{oreshkin2020nbeats}.

\emph{TFT} (Temporal Fusion Transformer) by \citet{lim2021temporal} is a forecasting architecture designed to combine the strengths of recurrent neural networks and transformer-based attention in a single, unified framework. The LSTM component processes historical prices sequentially to capture local patterns, while the attention mechanism identifies which past time steps are most relevant for forecasting -- automatically learning to weight recent observations more heavily during volatile periods or distant observations when long-term trends dominate. For multi-step forecasting, TFT outputs all $h$ future predictions simultaneously through a transformer-style attention decoder that attends to the encoded history (by the LSTM). Variable selection networks automatically identify which input features matter most, and the model provides interpretable attention weights showing which historical periods influenced each forecast.

We implement TFT with a simplified architecture (smaller hidden size and fewer attention heads) to accommodate the limited training data available for agricultural prices. The full architecture is extremely complex (500K+ parameters), requires massive training data, has many hyperparameters to tune, and trains slowly. For complete architectural details, see \citet{lim2021temporal}.

%====================================================================
\subsubsection{Foundation Models}

Foundation models represent a fundamentally different approach from the deep learning models above: rather than training from scratch on limited agricultural data, they leverage massive pre-training on diverse time series to learn general temporal patterns that transfer to new domains. We use publicly available pre-trained models developed by major technology firms including Amazon (Chronos, Chronos-2), Google (TimesFM 2.5), Salesforce (Moirai-2), and academic institutions (Time-MoE). Pre-training means the model has already learned patterns from millions of time series across diverse domains (retail sales, web traffic, energy consumption, financial data, etc.).\footnote{Foundation models are pre-trained on millions of time series from diverse domains, raising the possibility that they might have seen and even memorized agricultural price or related series during training. To address potential data leakage concerns, we conduct simulation analysis in Appendix~\ref{sec:app_simulation}, evaluating foundation models on synthetically generated price series that share the statistical properties of commodity prices but could not have appeared in any training data. The results confirm that strong performance reflects learned temporal patterns rather than memorization of specific agricultural series.} We evaluate the models in zero-shot mode, where we can directly apply their pre-trained weights without any fine-tuning on our specific agricultural data. This enables them to recognize common temporal patterns -- trends, seasonality, volatility -- without requiring extensive domain-specific training.\footnote{Unlike models above where we discuss architectural details, we focus here on the key features and distinguishing characteristics of each foundation model. The foundation model literature is rapidly evolving and architecturally complex; readers interested in implementation details should consult the original papers cited for each model.} Most foundation models generate multi-step forecasts autoregressively (predicting one value at a time), though Moirai-2 uses multi-token prediction to forecast multiple values simultaneously for improved efficiency.

\emph{Chronos} by \citet{ansari2024chronos} takes a novel approach by treating time series forecasting as a language modeling problem. The key insight is that language models excel at predicting the next word in a sequence -- Chronos applies this same principle to predict the next value in a time series. The model converts numerical prices into discrete tokens (similar to how text is converted to word tokens),\footnote{Tokenization is a fundamental concept in natural language processing where continuous text is broken into discrete units (words or subwords) that models can process. For example, the sentence ``The price is \$5.50'' might be tokenized as [``The'', ``price'', ``is'', ``\$'', ``5'', ``.'', ``50'']. Chronos applies this principle to numerical time series through a two-step process: first, it scales the time series by its absolute mean; second, it quantizes the scaled values into a fixed number of uniformly spaced bins. For instance, a scaled price of 1.23 might be mapped to bin 127 (token 127), representing values in the range [1.20, 1.25]. This discretization allows transformers designed for discrete sequences to process continuous numerical data. Other time series foundation models use similar tokenization strategies, though some work directly with continuous embeddings rather than discrete bins. See \citet{vaswani2017attention} for transformer architectures and \citet{devlin2019bert} for tokenization in language models.}  then uses a transformer architecture to predict future tokens, which are decoded back to price values. Specifically, Chronos builds on T5, a text-to-text transformer originally developed for natural language tasks. The model contains 200 million parameters and was pre-trained on over 100 diverse time series datasets. We use the \texttt{amazon/chronos-t5-base} model, generating 20 sample paths and taking the median for point forecasts. Chronos offers zero-shot capability, requires no hyperparameter tuning, and handles variable-length inputs. The tokenization process may lose some numerical precision compared to models that work directly with continuous values.

\emph{Chronos-2} \citep{ansari2025chronos2} is an updated version released in October 2025 that extends the original Chronos model with several architectural improvements. The key innovation is a mixture-of-experts (MoE) architecture. Rather than relying on a single monolithic network, the model consists of multiple specialized “expert” sub-networks, each trained to capture distinct patterns in agricultural prices such as stable trends, seasonal dynamics, or volatile periods. A learned gating mechanism allows the data to determine which experts are most relevant for a given input by assigning expert-specific weights based on current conditions -- for example, emphasizing one expert during trending periods and another during pronounced seasonal fluctuations. This specialization enables the model to achieve strong predictive performance with fewer total parameters (120 million vs. 200 million in the original). Chronos-2 also benefits from improved pre-training procedures and extends from univariate to universal forecasting, meaning it can handle multiple related time series simultaneously. We use \texttt{amazon/chronos-2} with the same inference settings as the original Chronos model. Interestingly, Chronos-2 underperforms the original Chronos in our experiments, demonstrating that newer models are not always better for specific applications.

%The key innovation here is a mixture-of-experts (MoE) design, where instead of using a single monolithic network, the model contains multiple specialized ``expert'' sub-networks, each expert trained to capture particular patterns (e.g., stable trends, seasonal dynamics, or volatile episodes). The model lets the data determine which expert should be used for a given situation, with a learned gating mechanism. The gating mechanism examines each input and routes it to the most appropriate experts -- for example, one expert might specialize in trending patterns while another handles seasonal fluctuations. This specialization allows the model to achieve strong performance with fewer total parameters (120 million vs. 200 million in the original). Chronos-2 also benefits from improved pre-training procedures and extends from univariate to universal forecasting, meaning it can handle multiple related time series simultaneously. We use \texttt{amazon/chronos-2} with the same inference settings as the original Chronos model. Interestingly, Chronos-2 underperforms the original Chronos in our experiments, demonstrating that newer models are not always better for specific applications.

\emph{TimesFM 2.5} (Time Series Foundation Model) by \citet{das2024timesfm} is a decoder-only transformer with 200 million parameters, pre-trained on an exceptionally large corpus of over 100 billion time points from diverse domains. Similar to how GPT-style language models work, the ``decoder-only'' architecture means the model processes the input sequence and generates outputs in a single forward pass (without a separate stage that first “encodes” the input into a fixed representation). Two key innovations distinguish TimesFM: first, it uses ``patching'' which groups consecutive time points into chunks before processing, allowing the model to capture local patterns efficiently while handling variable-length inputs. Second, it incorporates frequency-aware positional encodings that help the model understand different temporal granularities -- whether the data are hourly, daily, or monthly -- without explicit specification. We use \texttt{google/timesfm-2.5-200m-pytorch} with maximum context length of 512 time steps, processing the most recent 512 months of history (or the full history if shorter). TimesFM's massive pre-training corpus gives it exposure to an exceptionally wide range of temporal patterns, particularly relevant for agricultural forecasting.

\emph{Time-MoE} by \citet{jin2024timemoe} also implements a mixture-of-experts architecture but differs from Chronos-2 in several key aspects. First, Time-MoE also uses a decoder-only architecture (like TimesFM) rather than an encoder-decoder design, processing inputs and generating forecasts in a single forward pass. Second, it was pre-trained on Time-300B, a massive dataset spanning over 300 billion time points across 9 domains -- substantially larger than other foundation models' training corpora. Third, it supports flexible context lengths up to 4096 timepoints, allowing it to leverage longer historical sequences when available. We use the 50 million parameter variant (\texttt{Maple728/TimeMoE-50M}) with default inference settings. Despite being one of the smallest foundation models in our evaluation in terms of the parameters activated for any given input, Time-MoE will be shown to deliver the best overall performance, suggesting that massive pre-training scale and architectural efficiency can outweigh raw parameter count.

\emph{Moirai-2} by \citet{aksu2025moirai2} distinguishes itself through two key innovations: quantile forecasting and multi-token prediction. Unlike other foundation models that predict point forecasts, Moirai-2 directly outputs probabilistic forecasts across multiple quantiles (e.g., 10th, 50th, 90th percentiles), providing uncertainty estimates without requiring multiple sampling passes. Multi-token prediction means the model predicts multiple future values simultaneously in each forward pass rather than one at a time, substantially improving inference speed. The model uses a decoder-only architecture with single-patch inputs and quantile loss. We use \texttt{Salesforce/moirai-2.0-R-small} with 14 million parameters, the smallest variant in the Moirai-2 family, pre-trained on 36 million time series. Moirai-2 is twice as fast and thirty times smaller than its predecessor Moirai 1.0-Large while achieving better performance, demonstrating that architectural simplicity and efficient prediction strategies can outweigh model size.

%====================================================================
\subsection{Monthly Price and MYA Price Forecasts}

We conduct two separate analyses in our evaluation. The first evaluates monthly forecasting performance across all models, measuring the average error across 12 individual monthly forecasts. This analysis reveals which models best capture month-to-month price dynamics. The second evaluates MYA forecasts against USDA benchmarks. As noted in Section~\ref{sec:data}, USDA forecasts MYA directly (one value per marketing year). To enable fair comparison, we forecast 12 monthly prices for the next marketing year and then construct MYA forecasts using marketing percentage weights as defined in Equation (\ref{eq:mya_price}). This approach allows us to evaluate performance at both granularities while ensuring comparability with USDA's operational method. When monthly forecasts are aggregated to MYA, errors tend to partially cancel as overforecasts in some months offset underforecasts in others, so MYA MAE is typically lower than monthly MAE.

%====================================================================
\subsection{Evaluation Metrics}

We evaluate forecasts at both monthly and MYA aggregation levels using three standard metrics for each commodity-split combination.\footnote{While most of selected models can generate probabilistic forecasts with uncertainty quantification, we focus exclusively on point forecasts to maintain comparability with USDA, which publishes single-value price projections without prediction intervals or densities.} Mean Absolute Error (MAE) measures average absolute deviation:
\begin{align*}
   \text{MAE} &= \frac{1}{12}\sum_{t=1}^{12} |y_t - \hat{y}_t|.
\end{align*}
Root Mean Squared Error (RMSE) penalizes large errors more heavily:
\begin{align*}
   \text{RMSE} &= \sqrt{\frac{1}{12}\sum_{t=1}^{12} (y_t - \hat{y}_t)^2}.
\end{align*}
Mean Absolute Percentage Error (MAPE) expresses errors as percentages:
\begin{align*}
   \text{MAPE} &= \frac{100}{12}\sum_{t=1}^{12} \left|\frac{y_t - \hat{y}_t}{y_t}\right|,
\end{align*}
where $t$ indexes the 12 months in the marketing year. For each commodity, results are averaged across all cross-validation splits.

%====================================================================
\subsection{Cross-Validation Design and Hyperparameter Selection}

We implement an expanding window block cross-validation strategy with 16 temporal splits per commodity. Each split consists of three components: a training set (minimum 10 years, expanding over time), a validation set for hyperparameter selection (fixed at 2 years), and a test set for final evaluation (1 year). The first split uses data from September 1997 through August 2007 for training, September 2007 through August 2009 for validation, and September 2009 through August 2010 for testing; the final split uses September 1997 through August 2022 for training, September 2022 through August 2024 for validation, and September 2024 through August 2025 for testing.\footnote{These dates illustrate the corn, soybean, and wheat marketing year (September--August). Cotton follows an August--July marketing year. In practice, all splits strictly align with each commodity's official marketing year.} The training set grows by one year with each successive split while validation and test windows slide forward, mimicking operational forecasting where models are retrained as new data becomes available. All reported metrics are computed on the test set.

For models with tunable hyperparameters,\footnote{ETS, Prophet, Random Forest, XGBoost, LSTM, N-BEATS, TFT, and DeepAR. Foundation models are evaluated zero-shot without hyperparameter tuning. SARIMA uses automatic parameter selection. Naive and seasonal naive have no tunable parameters.} we conduct grid search: train candidate models on the training set, evaluate on the validation set, select the configuration with lowest validation RMSE, then retrain on combined training and validation data before generating test forecasts. The test year is never used for model selection decisions. We deliberately use focused grid searches (10--20 combinations) rather than exhaustive searches, as prior research shows extensive tuning can cause overfitting to small validation sets \citep{makridakis2018statistical}.

With 4 commodities and 16 splits, we generate 64 split-commodity combinations for model comparison and 1,088 total forecasts across all models ($17 \times 64$). For MYA price prediction comparison, data availability varies by commodity, yielding 49 USDA-comparable forecasts. To ensure fair comparison with USDA ERS, we use their second-to-last weekly-updated forecast before the marketing year begins to avoid any overlap with the marketing year. Our models forecast 12 monthly prices using only information available before the marketing year starts, then aggregate to SAP using the same marketing percentage weights. This ensures our models operate under a restrictive information set, with no future information leakage. See Appendix~\ref{sec:app_cv} for complete cross-validation details including split definitions, hyperparameter specifications, and optimization experiments.

%====================================================================
\section{Empirical Results}
\label{sec:results}

We evaluate model performance in two complementary steps: monthly price forecasting across all 17 models, and MYA forecasts benchmarked against USDA ERS SAP prediction. This section presents results for both sets of evaluations, identifying top performers and analyzing patterns across commodities.

%====================================================================
\subsection{Evaluation 1: Monthly Price Forecasting}

We evaluate all 17 models on 12-month-ahead forecasts using 64 cross-validation splits (16 splits × 4 commodities). For each split, we compute MAE, RMSE, and MAPE. Overall rankings average metrics across all splits; we also report commodity-specific performance. Models are ranked by MAE.

%====================================================================
\subsubsection{Overall Model Rankings}

Table \ref{tab:monthly_rankings} presents overall monthly forecasting performance across all commodities and splits. Foundation models dominate the top positions, with Time-MoE achieving the best performance.

\begin{table}[ht!]
   \centering
   \caption{Monthly Price Forecasting Performance - All Models}
   \label{tab:monthly_rankings}
   \footnotesize
   \begin{tabular}{@{}lccccr@{}}
   \toprule
   \textbf{Rank} & \textbf{Model} & \textbf{MAE} & \textbf{RMSE} & \textbf{MAPE (\%)} & \textbf{Category} \\
   \midrule
   1 & Time-MoE & 0.693 & 0.784 & 12.88 & Foundation \\
   2 & Chronos & 0.734 & 0.819 & 14.31 & Foundation \\
   3 & Chronos-2 & 0.736 & 0.819 & 13.86 & Foundation \\
   4 & TimesFM 2.5 & 0.736 & 0.817 & 13.86 & Foundation \\
   5 & Moirai-2 & 0.751 & 0.838 & 14.21 & Foundation \\
   6 & Naive & 0.775 & 0.849 & 14.38 & Traditional \\
   7 & Random Forest & 0.793 & 0.876 & 15.45 & Machine Learning \\
   8 & XGBoost & 0.823 & 0.918 & 15.80 & Machine Learning \\
   9 & SARIMA & 0.859 & 0.944 & 16.28 & Traditional \\
   10 & LSTM & 0.893 & 1.010 & 16.78 & Deep Learning \\
   11 & N-BEATS & 0.894 & 0.976 & 16.91 & Deep Learning \\
   12 & Exp Smoothing & 0.921 & 1.015 & 15.96 & Traditional \\
   13 & TFT & 0.961 & 1.044 & 17.19 & Deep Learning \\
   14 & Seasonal Naive & 0.973 & 1.071 & 17.13 & Traditional \\
   15 & STL & 1.033 & 1.150 & 18.41 & Traditional \\
   16 & DeepAR & 1.267 & 1.342 & 22.08 & Deep Learning \\
   17 & Prophet & 1.291 & 1.374 & 22.60 & Traditional \\
   \bottomrule
   \end{tabular}
   \vskip 2mm
   \begin{minipage}{0.98\textwidth}
      {\footnotesize \textit{Notes}: Monthly forecast performance averaged across 768 forecasts ($64$ splits $\times$ $12$ months, with some variation by commodity).\par}
   \end{minipage}
\end{table}

Several key findings emerge from the monthly forecasting results. Foundation models achieve superior performance, with all five ranking in the top five positions. Time-MoE leads with MAE \$0.693, followed closely by Chronos (\$0.734) and Chronos-2 (\$0.736). The zero-shot capability of pre-trained models proves highly effective for agricultural price forecasting.

Deep learning models trained from scratch consistently underperform, ranking 10th-16th out of 17 models. LSTM (rank 10, MAE \$0.893), N-BEATS (rank 11, MAE \$0.894), TFT (rank 13, MAE \$0.961), and DeepAR (rank 16, MAE \$1.267) all fail despite extensive hyperparameter optimization. With only 100-250 training samples and 50K-500K parameters, these models severely overfit.

Simple baselines remain competitive. The Naive model (rank 6, MAE \$0.775) outperforms all deep learning models and most traditional methods, demonstrating that simpler models generalize better with limited data. Random Forest (rank 7) and XGBoost (rank 8) also perform well, benefiting from ensemble methods that reduce overfitting.

%====================================================================
\subsubsection{Performance by Commodity}

Table \ref{tab:monthly_by_commodity} reveals commodity-specific patterns in model performance. No single model dominates across all commodities: Time-MoE leads on corn and wheat, Chronos on soybeans, and Seasonal Naive on cotton. This heterogeneity suggests that optimal model choice depends on commodity characteristics. Nevertheless, foundation models consistently rank in the top tier across all commodities, while deep learning models trained from scratch occupy the bottom tier despite extensive hyperparameter tuning.

Forecasting difficulty varies substantially by commodity. Cotton presents the easiest challenge with the narrowest performance spread (1.5× range), where even simple Seasonal Naive achieves competitive accuracy. Soybeans prove most challenging with the widest spread (2.1× range), reflecting high price volatility and complex market dynamics. Corn and wheat show intermediate difficulty with moderate differentiation (2.2× and 1.7× ranges respectively). The consistent ranking of foundation models above deep learning models across all commodities demonstrates that the performance gap reflects fundamental differences in model architectures and training approaches rather than commodity-specific factors.

\begin{table}[htp!]
   \centering
   \caption{Monthly Forecasting Performance by Commodity (MAE, \$/unit)}
   \label{tab:monthly_by_commodity}
   \footnotesize
   \begin{tabular}{@{}lcccc@{}}
   \toprule
   \textbf{Model} & \textbf{Corn} & \textbf{Soybeans} & \textbf{Wheat} & \textbf{Cotton} \\
   \midrule
   \multicolumn{5}{l}{\textit{Foundation Models (Zero-Shot)}} \\
   Time-MoE & \textbf{0.638} & 1.192 & \textbf{0.828} & 0.115 \\
   TimesFM 2.5 & 0.668 & 1.206 & 0.947 & 0.122 \\
   Chronos-2 & 0.677 & 1.135 & 1.015 & 0.117 \\
   Chronos & 0.703 & \textbf{1.105} & 0.992 & 0.135 \\
   Moirai-2 & 0.661 & 1.206 & 1.018 & 0.121 \\
   \midrule
   \multicolumn{5}{l}{\textit{Traditional Time Series}} \\
   Naive & 0.712 & 1.247 & 1.025 & 0.117 \\
   Seasonal Naive & 0.929 & 1.579 & 1.275 & \textbf{0.109} \\
   SARIMA & 0.868 & 1.273 & 1.172 & 0.123 \\
   Exp Smoothing & 0.934 & 1.553 & 1.081 & 0.115 \\
   STL & 1.078 & 1.807 & 1.122 & 0.126 \\
   Prophet & 1.295 & 2.388 & 1.359 & 0.121 \\
   \midrule
   \multicolumn{5}{l}{\textit{Machine Learning}} \\
   Random Forest & 0.669 & 1.273 & 1.091 & 0.139 \\
   XGBoost & 0.650 & 1.425 & 1.078 & 0.138 \\
   \midrule
   \multicolumn{5}{l}{\textit{Deep Learning}} \\
   N-BEATS & 0.868 & 1.471 & 1.108 & 0.129 \\
   LSTM & 0.908 & 1.491 & 1.058 & 0.115 \\
   TFT & 0.964 & 1.553 & 1.212 & 0.115 \\
   DeepAR & 1.204 & 2.241 & 1.459 & 0.165 \\
   \bottomrule
   \end{tabular}
   \vskip 2mm
   \begin{minipage}{0.98\textwidth}
      {\footnotesize \textit{Notes}: Bold values indicate best performance for each commodity.\par}
   \end{minipage}
\end{table}

%====================================================================
\subsubsection{Statistical Significance of Performance Differences}

To formally test whether the observed performance differences are statistically significant, we conduct pairwise Diebold-Mariano (DM) tests across all model pairs. Unlike settings with a single established benchmark, our evaluation spans four model categories (traditional, ML, DL, foundation) with no clear reference model. To comprehensively assess performance differences both within and across categories, we test all pairwise comparisons. Following \citet{diebold1995comparing}, we test the null hypothesis of equal predictive accuracy using absolute percentage errors (APE) as the loss differential, which provides scale-invariance when pooling forecasts across commodities with different price levels. We compute Newey-West HAC standard errors to account for serial correlation in forecast errors.

\begin{table}[ht!]
   \centering
   \caption{Diebold-Mariano Test Results: TSFM Win Rates Against Baseline Models}
   \label{tab:dm_winrate}
   \begin{tabular}{lcc}
      \toprule
      & \multicolumn{2}{c}{TSFM Wins} \\
      \cmidrule(lr){2-3}
      Comparison & Total & $p<0.05$ \\
      \midrule
      vs Traditional & 29/30 & 21/30 \\
      vs ML & 10/10 & 8/10 \\
      vs DL & 20/20 & 20/20 \\
      \midrule
      Total & 59/60 & 49/60 \\
      \bottomrule
   \end{tabular}
   \vskip 2mm
   \begin{minipage}{0.98\textwidth}
      {\footnotesize \textit{Notes:} Diebold-Mariano tests compare each TSFM (5 models) against baseline models using absolute percentage errors pooled across all commodities and forecast horizons. Traditional category includes Naive, Seasonal Naive, SARIMA, Exponential Smoothing, STL, and Prophet (30 comparisons). ML category includes Random Forest and XGBoost (10 comparisons). DL category includes LSTM, N-BEATS, TFT, and DeepAR (20 comparisons). Positive DM statistics indicate TSFM outperformance. Full pairwise results in Appendix \ref{sec:app_dm_tests}.\par}
   \end{minipage}
\end{table}

Table \ref{tab:dm_winrate} summarizes DM test results for foundation models against all other models. TSFMs win 59 of 60 comparisons, with 49 statistically significant at the 5\% level. The dominance is most pronounced against deep learning models, where all 20 comparisons favor TSFMs significantly. The single non-win is Chronos versus Naive, where Chronos achieves lower average errors but the difference is not statistically significant. These results confirm that TSFM performance advantages are statistically robust.

To verify that foundation models learn genuine forecasting patterns rather than memorizing historical sequences, we evaluate all 17 models on 400 synthetically generated price series matching the statistical properties of agricultural data (Appendix \ref{sec:app_simulation}). Foundation models maintain top performance on synthetic data, with Chronos-2, Moirai-2, and TimesFM ranking 1-3, confirming they have learned generalizable time series patterns rather than memorizing specific historical sequences. This validation addresses concerns about potential data contamination in foundation model pre-training corpora.

%====================================================================
\subsection{Evaluation 2: MYA Forecasting vs USDA Benchmark}

We evaluate MYA forecasts against USDA ERS forecast, which is an important input to the prediction in WASDE report. ERS forecast represent a rather competitive benchmark to compare with for two fundamental reasons. First, USDA's futures-plus-basis methodology incorporates forward-looking market information unavailable to our models. Futures prices aggregate expectations from thousands of market participants regarding supply conditions, demand shifts, weather patterns, trade policy developments, and other fundamental factors \citep{fama1970efficient, tomek1997commodity}. This information set extends well beyond the historical price patterns our models observe. Second, USDA forecasts each month directly using the corresponding nearby futures contract (December futures for October-November prices, March futures for December-February prices, and so forth) rather than iteratively. This direct approach avoids the error accumulation inherent in sequential forecasting across the 12-month horizon. These dual advantages (superior information and methodology) establish USDA as a particularly demanding benchmark. Consequently, models that outperform USDA demonstrate an ability to extract predictive patterns from historical prices that futures markets do not fully incorporate.

Our evaluation spans marketing years (MY) 2017-2024 excluding the COVID shock year (2020), yielding 7 years of data for all commodities to ensure balanced comparison.\footnote{We focus on MY 2017-2024 for two reasons. First, this period provides a balanced panel where all four commodities have complete data, enabling fair cross-commodity comparison. Cotton MYA forecasts only became available from MY 2017 onward. Second, we exclude MY 2020 (September 2020 - August 2021) because the COVID-19 pandemic created unprecedented market disruptions that distort forecast errors. Including this outlier year would mask performance patterns in normal market conditions. Our focus on MY 2017-2024 is driven by data availability constraints rather than model performance considerations. This period includes substantial market volatility (2018-2019 U.S.-China trade war, 2019 Midwest flooding, 2022-2023 Ukraine war impacts), providing a stringent test of model robustness across diverse market conditions. While longer evaluation periods would strengthen external validity, extending backward would require either dropping cotton or analyzing an unbalanced panel, complicating cross-commodity comparisons.} We use USDA's second-to-last forecast before each marketing year starts to eliminate ambiguity regarding week boundaries. Models aggregate their 12-month forecasts to MYA using 5-year averaged marketing percentage weights (7-year Olympic average for cotton), replicating USDA methodology detailed in Section \ref{sec:mya_prices}. For each model, we compute commodity-specific MAE by averaging absolute errors across matched years.

%====================================================================
\subsubsection{Best Model Performance by Commodity}

Table \ref{tab:best_by_commodity} presents the best-performing model for each commodity. Time-MoE achieves substantial improvements on corn (18.5\%) and wheat (54.9\%), while TimesFM performs best on soybeans (6.9\%). Cotton remains challenging, with USDA maintaining superior accuracy.

\begin{table}[ht!]
   \centering
   \caption{Best Model Performance vs USDA by Commodity (MY 2017-2024, excl 2020)}
   \label{tab:best_by_commodity}
   \footnotesize
   \begin{tabular}{@{}lcccc@{}}
      \toprule
      \textbf{Commodity} & \textbf{USDA MAE} & \textbf{Best Model} & \textbf{Model MAE} & \textbf{Improvement} \\
      \midrule
      Corn & 0.328 & Time-MoE & 0.267 & +18.5\% \\
      Soybeans & 0.396 & TimesFM & 0.369 & +6.9\% \\
      Wheat & 1.201 & Time-MoE & 0.542 & +54.9\% \\
      Cotton & 0.045 & Seasonal Naive & 0.094 & $-$110.5\% \\
      % \midrule
      % \textbf{Overall} & 0.492 & \textbf{Time-MoE} & 0.343 & +30.3\% \\
      \bottomrule
   \end{tabular}
   \vskip 2mm
   \begin{minipage}{0.98\textwidth}
      {\footnotesize \textit{Notes}: Best model selected based on lowest MAE for each commodity. Improvement = (USDA MAE - Model MAE) / USDA MAE. Negative improvement indicates USDA outperforms the model.\par}
   \end{minipage}
\end{table}

% Figure \ref{fig:usda_comparison} illustrates these commodity-specific performance differences. Panel (a) shows absolute MAE values for USDA versus the best model for each commodity. Panel (b) displays percentage improvements, highlighting the substantial gains on major row crops and more modest improvements on wheat and cotton.

% \begin{figure}[ht!]
%    \centering
%    \caption{Best Model Performance vs USDA by Commodity}
%    \label{fig:usda_comparison}
%    \begin{subfigure}[b]{0.48\textwidth}
%       \centering
%       \includegraphics[width=\textwidth]{figures/figure3a_mae_comparison.png}
%       \caption{MAE Comparison}
%       \label{fig:usda_mae}
%    \end{subfigure}
%    \hfill
%    \begin{subfigure}[b]{0.48\textwidth}
%       \centering
%       \includegraphics[width=\textwidth]{figures/figure3b_improvement.png}
%       \caption{Improvement over USDA}
%       \label{fig:usda_improvement}
%    \end{subfigure}
%    \vskip 2mm
%    \begin{minipage}{0.98\textwidth}
%       {\footnotesize \textit{Notes:} Panel (a) shows Marketing Year Average (MYA) forecast MAE for USDA ERS operational forecasts versus the best-performing model for each commodity. Panel (b) shows percentage improvement of the best model over USDA. XGBoost achieves the largest improvement on corn (+55.9\%), while Time-MoE performs best on soybeans (+53.8\%), wheat (+21.5\%), and cotton (+29.0\%).\par}
%    \end{minipage}
% \end{figure}

%====================================================================
\subsubsection{Comprehensive Model Comparison Across Commodities}

Table \ref{tab:all_models_by_commodity} presents MYA forecast performance for all 17 models across four commodities, expressed as ratios relative to USDA ERS (values below 1.00 indicate superior performance). Foundation models demonstrate strong performance, with Time-MoE achieving the best results on corn (0.82) and wheat (0.45). TimesFM performs best on soybeans (0.93), representing the only model across all categories to outperform USDA on this commodity. Traditional and machine learning models show mixed results, with several achieving competitive performance on wheat. Deep learning models trained from scratch show strong performance on wheat, with LSTM achieving 0.55 (45\% improvement), though they underperform on other commodities despite extensive hyperparameter optimization.

\begin{table}[ht!]
   \centering
   \caption{Model Performance by Commodity - Relative to USDA ERS}
   \label{tab:all_models_by_commodity}
   \footnotesize
   \begin{tabular}{lllll}
      \toprule
      \textbf{Model} & \textbf{Corn} & \textbf{Soybeans} & \textbf{Wheat} & \textbf{Cotton} \\
      \midrule
      \textbf{USDA ERS (Baseline)} & 1.00 & 1.00 & 1.00 & 1.00 \\
      \midrule
      \multicolumn{5}{l}{\textit{Foundation Models (Zero-Shot)}} \\
      Time-MoE & \textbf{0.82} & 1.18 & \textbf{0.45}$^{**}$ & 2.15 \\
      TimesFM & 1.08 & \textbf{0.93} & \textbf{0.63}$^{**}$ & 2.99 \\
      Chronos & \textbf{0.97} & 1.18 & \textbf{0.69}$^{*}$ & 3.41 \\
      Chronos-2 & 1.31 & 1.01 & \textbf{0.65}$^{**}$ & 2.77 \\
      Moirai-2 & 1.14 & 1.21 & \textbf{0.68}$^{**}$ & 2.90 \\
      \midrule
      \multicolumn{5}{l}{\textit{Machine Learning}} \\
      Random Forest & 1.20 & 1.48 & \textbf{0.61}$^{*}$ & 3.09 \\
      XGBoost & 1.17 & 1.28 & \textbf{0.58} & 2.87 \\
      \midrule
      \multicolumn{5}{l}{\textit{Deep Learning}} \\
      LSTM & 1.65 & 1.63 & \textbf{0.55} & 2.39 \\
      N-BEATS & 1.52 & 1.72 & \textbf{0.64}$^{**}$ & 3.24 \\
      TFT & 2.17 & 2.09 & \textbf{0.99} & 2.11 \\
      DeepAR & 2.23 & 4.23 & \textbf{0.94} & 3.44 \\
      \midrule
      \multicolumn{5}{l}{\textit{Traditional Time Series}} \\
      Naive & 1.44 & 1.46 & \textbf{0.73}$^{*}$ & 3.19 \\
      Seasonal Naive & 1.95 & 2.99 & 1.05 & 2.11 \\
      SARIMA & 1.65 & 1.71 & \textbf{0.76}$^{*}$ & 2.91 \\
      Exp Smoothing & 2.54 & 1.66 & \textbf{0.82} & 2.98 \\
      STL & 2.61 & 2.96 & 1.02 & 2.46 \\
      Prophet & 3.72 & 5.13 & 1.19 & 2.71 \\
      \bottomrule
   \end{tabular}
   \vskip 2mm
   \begin{minipage}{0.98\textwidth}
      {\footnotesize \textit{Notes:} Values show model MAE relative to USDA ERS baseline (model MAE / ERS MAE). Bold values indicate model outperforms ERS (ratio $<$ 1.00). Significance stars from Diebold-Mariano tests: $^{*}$ p$<$0.10, $^{**}$ p$<$0.05, $^{***}$ p$<$0.01. Stars appear only when model outperforms ERS. DM tests use Newey-West HAC standard errors with h=1 lag (annual MYA errors are non-overlapping across marketing years). USDA forecasts use second-to-last forecast before marketing year starts. All models evaluated using same test years for fair comparison.\par}
   \end{minipage}
\end{table}

The results reveal substantial heterogeneity across commodities. Wheat exhibits the strongest performance, with 14 out of 17 models outperforming USDA, led by Time-MoE at 0.45 (55\% improvement). Corn shows modest improvements, with Time-MoE (0.82) and Chronos (0.97) beating USDA. Soybeans remains competitive, with only TimesFM (0.93) outperforming USDA. Cotton proves most challenging, with all models substantially worse than USDA (ratios 2.11-3.44), reflecting the exceptional accuracy of USDA's futures-based forecasts for this commodity. These results are particularly noteworthy given that our models use only historical prices while USDA incorporates forward-looking futures market information. The fact that pure time series models can outperform futures-based benchmarks on major commodities demonstrates their ability to extract predictive patterns not fully reflected in market expectations.

Our evaluation focuses on pre-season forecasts generated before the marketing year begins, with matched USDA ERS forecasts. This design choice reflects practical forecasting constraints: policymakers and farmers require advance price projections to inform planting decisions and program enrollment. Research on fixed-event forecasts documents that accuracy varies substantially by forecast horizon, with errors declining as more information becomes available closer to the target period \citep{isengildina2024optimal}. Our single-horizon design provides policy-relevant comparisons at the most challenging forecast timing, though future work could examine whether foundation model advantages persist across multiple horizons.

\subsection{Data Leakage Evaluation Framework}

A critical concern in evaluating Time Series Foundation Models (TSFMs) is distinguishing between genuine pattern learning and memorization of training data. As \citet{meyer2025tsfm} highlight, the field faces significant challenges from "risks of information leakage due to overlapping and obscure datasets" and "memorization of global patterns," yet lacks established evaluation methodologies to address these concerns. Appendix (\ref{sec:app_simulation}) presents a comprehensive framework for evaluating data leakage risks through multiple complementary approaches, and our conclusions remain the same.

%====================================================================
\section{Discussion}
\label{sec:discussion}

%====================================================================
\subsection{Practical Implications for Agricultural Policy}

WASDE SAP forecasts play a central role in administering Farm Bill safety net programs, where Price Loss Coverage (PLC) and Agriculture Risk Coverage (ARC) payments depend on MYA price realizations \citep{hoffman2018cotton}. Time-MoE achieves strong performance relative to USDA ERS forecasts on recent data, with particularly notable results on wheat (55\% improvement) and corn (18\% improvement) -- the two largest U.S. field crops representing over 175 million acres annually. These improvements could translate to more reliable program payment estimates, potentially affecting billions of dollars in farm safety net expenditures.

Operational deployment by USDA would require addressing explainability concerns, as foundation models operate as black boxes. A hybrid approach could use foundation models to generate primary forecasts while maintaining traditional methods (SARIMA, futures-basis) to provide interpretable validation checks and explain forecast drivers to policy stakeholders. For operational implementation, USDA could generate foundation model forecasts as complementary inputs to their existing futures-based methodology. When the two approaches diverge substantially, this signals either unusual market conditions or potential basis forecast errors, triggering additional analyst review. The modest computational requirements (sub-second inference after one-time model download) make this operationally feasible without significant infrastructure investment.

Beyond farm policy applications, our findings have broader implications for commodity market participants. Previous research demonstrates that futures-based forecasts provide viable alternatives to USDA forecasts for private sector users \citep{colino2010, hoffman2015, irwin1994, irwin2015, manfredo2004, poghosyan2025cotton}. Our results suggest that foundation models, requiring only historical prices and no specialized market data or futures market access, offer an accessible forecasting tool for agricultural businesses, commodity traders, and financial institutions. Companies using commodities for portfolio diversification or hedging strategies can leverage these models to generate reliable price projections without the data infrastructure and market access required for futures-based approaches.

It is worth noting that foundation model advantages vary by commodity: strong for wheat and corn, modest for soybeans, absent for cotton. This suggests commodity-specific model selection rather than universal application.

%====================================================================
\subsection{Why Deep Learning Models Fail on Limited Agricultural Data}

Deep learning models trained from scratch (LSTM, N-BEATS, TFT, DeepAR) consistently underperform despite extensive hyperparameter optimization, ranking 10th-16th out of 17 models. Agricultural price data provides only 120-250 training samples, while these models have 50,000-500,000 parameters -- yielding parameter-to-sample ratios of 200:1 to 4,000:1. \citet{goodfellow2016deep} recommend minimum 5,000 samples for acceptable deep learning performance; our data provides only 2-5\% of this threshold. This fundamental data scarcity explains why complex neural architectures consistently overfit, memorizing training patterns rather than learning generalizable features.

We conducted grid search experiments to test whether more extensive hyperparameter tuning could improve performance. The focused grid searched over core parameters: n\_lags, n\_estimators, and max\_depth for Random Forest; n\_lags, n\_estimators, and learning\_rate for XGBoost. The enhanced grid added regularization and sampling parameters: learning rate and subsample for Random Forest; max\_depth, subsample, and colsample\_bytree for XGBoost. Counterintuitively, models selected from the enhanced grid performed worse than those from the focused grid. This suggests that with limited validation data (2 years = 24 months), expanding the hyperparameter search space increases the risk of overfitting to validation set idiosyncrasies rather than identifying genuinely better configurations.

Foundation models avoid this problem entirely through pre-training on millions of time series. Their parameters are already learned from diverse domains; zero-shot inference requires no agricultural-specific training, eliminating overfitting. This explains why Time-MoE with 50 million parameters outperforms LSTM with 50 thousand parameters -- the former leverages knowledge from massive pre-training, while the latter must learn everything from 200 agricultural samples.

%====================================================================
\subsection{Why Foundation Models Outperform Futures-Based Forecasts}

Foundation models achieve substantial improvements over USDA ERS forecasts on wheat (55\%) and corn (18\%), modest gains on soybeans (7\%), but underperform on cotton. This pattern is particularly striking given that our models use only historical cash prices while USDA incorporates forward-looking futures prices from actively traded markets. Under market efficiency, futures prices should aggregate all available information, making them difficult to outperform. Two explanations could account for foundation model success: either they capture price dynamics not fully reflected in futures markets, or systematic errors in USDA's basis adjustment methodology create exploitable forecast biases.

The latter explanation appears more plausible. USDA's futures-plus-basis methodology assumes that historical basis patterns (5-7 year averages) reliably predict future basis realizations. However, basis reflects local market conditions -- storage costs, transportation infrastructure, regional supply-demand imbalances -- that evolve over time. Infrastructure improvements, policy reforms, market consolidation, and logistics technology changes can alter basis patterns, making historical averages poor predictors during periods of structural change.

Empirical evidence supports this interpretation. \citet{hoffman2015} found that WASDE corn price projections outperformed futures-adjusted forecasts in only 4 out of 16 forecasting periods, suggesting substantial room for improvement. \citet{isengildina2024optimal} document systematic underestimation bias in USDA price forecasts for corn, soybeans, and wheat, likely reflecting analysts' tendency to underestimate long-term growth rates. \citet{poghosyan2025cotton} demonstrate that incorporating current basis deviations from historical averages significantly improves cotton price forecast accuracy, confirming that fixed historical basis averages miss important market information. Foundation models may exploit similar patterns by learning adaptive relationships directly from cash price dynamics rather than relying on fixed historical basis adjustments. This mechanism explains why foundation models excel on wheat and corn, where basis patterns have experienced more structural change, while struggling on cotton, where USDA's 7-year Olympic average basis methodology proves more stable and accurate.

%====================================================================
\subsection{Univariate Forecasting Design}

Our evaluation focuses on purely univariate time series forecasting -- each model uses only the historical price series of a single commodity to generate forecasts. This design choice is deliberate and reflects standard practice in applied forecasting, where univariate methods serve as the primary benchmark for evaluating model performance \citep{makridakis2018m4, hyndman2018forecasting}. Univariate forecasting isolates the contribution of model architecture and temporal pattern recognition from the confounding effects of feature engineering and covariate selection.

We recognize that several extensions could potentially improve forecast accuracy: pooling data across commodities, incorporating exogenous covariates (e.g., futures prices, weather, macroeconomic indicators), or multivariate forecasting that captures cross-commodity dynamics. These represent planned next steps in our research agenda. However, our objective in this study is to establish a fair comparison of forecasting methods under identical information constraints -- using only each commodity's own price history. This ``pure'' univariate setting provides a clean test of whether foundation models can extract predictive patterns from limited historical data, which is the fundamental question motivating our research.

The strong performance of foundation models in this univariate setting is particularly noteworthy: they achieve substantial improvements over USDA benchmarks despite using less information (no futures prices or basis). This suggests that the patterns learned from diverse pre-training datasets transfer effectively to agricultural price forecasting, even without domain-specific covariates or cross-commodity pooling.

%====================================================================
\subsection{Limitations and Future Research}

Our evaluation focuses on four major U.S. crops at monthly frequency using purely univariate forecasting. While this design provides a clean test of foundation model capabilities under controlled conditions, several extensions could strengthen external validity and practical applicability.

\textit{Data scope.} Our sample focuses on four major field crops with liquid futures markets and established USDA forecasting benchmarks. The broader USDA forecasting system encompasses livestock production (beef, pork, poultry), dairy prices, and specialty crops, each presenting distinct forecasting challenges \citep{isengildina2024optimal}. Extending our evaluation to livestock and dairy would test foundation model performance on commodities with different biological production cycles (gestation periods vs. growing seasons) and storage constraints (perishability vs. storability). Specialty crops (fruits, vegetables, nuts) present additional challenges: limited or no futures markets, higher price volatility from weather sensitivity, and stronger regional variation in production and pricing. International commodity prices from FAO or World Bank databases would assess whether foundation models generalize across different policy regimes, trade patterns, and market structures. Higher-frequency data (daily futures prices) would test short-horizon forecasting capabilities relevant for trading and risk management.

\textit{Methodological scope.} Our univariate approach treats each commodity independently. Cross-commodity pooling could increase training sample size by combining data from related commodities (e.g., training a single model on corn, soybeans, and wheat together), though this requires addressing different price scales and seasonality patterns. Multivariate forecasting could explicitly model interdependencies: corn and soybean prices are linked through crop rotation decisions, land allocation, and shared input costs (fertilizer, fuel); wheat and corn compete in livestock feed markets, creating substitution effects. Covariate-augmented models could test whether foundation models effectively integrate exogenous information such as futures price curves (forward-looking market expectations), weather indices (drought monitors, temperature anomalies), and macroeconomic indicators (exchange rates, energy prices). Probabilistic forecasting with prediction intervals would provide uncertainty quantification for risk assessment in farm program administration.

\textit{Related applications.} Beyond price forecasting, foundation models could address other agricultural prediction problems. Crop yield forecasting combines weather patterns, soil conditions, and agronomic practices to predict production outcomes. Supply-demand balance projections require integrating production forecasts with consumption trends, inventory dynamics, and trade flows. International trade flow forecasting involves complex interactions between domestic production, foreign demand, exchange rates, and trade policies. Each application presents unique data structures and domain constraints that would test foundation model adaptability across agricultural economics.

%====================================================================
\section{Conclusion}
\label{sec:conclusion}

This paper provides the first systematic evaluation of time series foundation models for agricultural price forecasting, comparing five foundation models against 12 baselines using USDA ERS data for corn, soybeans, wheat, and cotton from 1997-2025.

For monthly price forecasting, foundation models dominate performance rankings, occupying all top five positions. Time-MoE achieves the lowest mean absolute error, followed by Chronos, Chronos-2, TimesFM 2.5, and Moirai-2. Deep learning models trained from scratch rank 10th-16th despite extensive hyperparameter optimization, while the Naive model (rank 6) outperforms all deep learning approaches. For MYA forecasting against USDA ERS benchmarks, foundation models achieve strong performance on major commodities despite using only historical prices while USDA incorporates forward-looking futures market information. On recent data, Time-MoE delivers substantial improvements on wheat and corn, while TimesFM achieves modest gains on soybeans. These findings challenge the long-standing belief that simple models forecast best in agricultural markets.

These findings have direct policy implications for Farm Bill program administration, where improved SAP forecasts could affect billions of dollars in payments to U.S. farmers. Beyond farm policy applications, our findings have implications for commodity market participants. Prior research notes that futures-based forecasts provide viable alternatives to USDA forecasts for private sector users. Our results suggest that foundation models, requiring only historical prices and no specialized market data, offer an accessible forecasting tool for agricultural businesses, commodity traders, and portfolio managers seeking reliable price projections.

More broadly, our findings have implications beyond agricultural forecasting. The data characteristics that challenge agricultural price prediction (limited samples, structural breaks, strong seasonality, high volatility) are common across economic and business forecasting domains. Foundation models' ability to outperform specialized benchmarks despite these challenges suggests their potential applicability to other data-scarce forecasting problems, including financial time series, macroeconomic indicators, and business demand forecasting. The synthetic benchmark results (Appendix \ref{sec:app_simulation}) further demonstrate that these advantages stem from genuine pattern learning rather than domain-specific memorization, as foundation models maintain top performance on synthetically generated data with zero contamination risk. Our results demonstrate that foundation models offer a viable path forward for economic forecasting in data-scarce domains where traditional machine learning has historically struggled.

\clearpage
\bibliography{tsfm_forecasting}

@article{meyer2025tsfm,
	title={Time Series Foundation Models: Benchmarking Challenges and Requirements},
	author={Meyer, M. and Kaltenpoth, S. and Zalipski, K. and M{\"u}ller, O.},
	journal={arXiv preprint arXiv:2510.13654},
	year={2025},
	url={https://arxiv.org/abs/2510.13654}
}

@article{ansari2024chronos,
	title={Chronos: Learning the Language of Time Series},
	author={Ansari, Abdul Fatir and Stella, Lorenzo and Turkmen, Caner and Zhang, Xiyuan and Mercado, Pedro and Shen, Huibin and Shchur, Oleksandr and Rangapuram, Syama Sundar and Pineda Arango, Sebastian and Kapoor, Shubham and others},
	journal={Transactions on Machine Learning Research},
	year={2024},
	url={https://openreview.net/forum?id=gerNCVqqtR}
}

@article{ansari2025chronos2,
	title={Chronos-2: From Univariate to Universal Forecasting},
	author={Ansari, Abdul Fatir and Shchur, Oleksandr and K{\"u}ken, Jaris and Auer, Andreas and Han, Boran and Mercado, Pedro and Rangapuram, Syama Sundar and Shen, Huibin and Stella, Lorenzo and Zhang, Xiyuan and others},
	journal={arXiv preprint arXiv:2510.15821},
	year={2025},
	url={https://arxiv.org/abs/2510.15821}
}

@inproceedings{jin2024timemoe,
	title={Time-MoE: Billion-Scale Time Series Foundation Models with Mixture of Experts},
	author={Jin, Ming and Wen, Qingsong and Liang, Yuxuan and Zhang, Chaoli and Xue, Siqiao and Wang, Xue and Zhang, James and Wang, Yi and Chen, Haifeng and Li, Xiaoli and others},
	booktitle={International Conference on Learning Representations (ICLR)},
	year={2025},
	url={https://openreview.net/forum?id=EvGGezLCHq}
}

@article{aksu2025moirai2,
	title={When Less Is More for Time Series Forecasting},
	author={Aksu, Taha and Woo, Gerald and Liu, Chenghao and Savarese, Silvio and Xiong, Caiming and Sahoo, Doyen},
	journal={arXiv preprint arXiv:2511.11698},
	year={2025},
	url={https://arxiv.org/abs/2511.11698}
}

@inproceedings{das2024timesfm,
	title={A Decoder-Only Foundation Model for Time-Series Forecasting},
	author={Das, Abhimanyu and Kong, Weihao and Sen, Rajat and Zhou, Yichen},
	booktitle={International Conference on Machine Learning (ICML)},
	pages={10148--10167},
	year={2024}
}

@article{zelingher2024agricaf,
	title={{AGRICAF}: A New Methodology for Agricultural Commodity Analysis and Forecasts},
	author={Zelingher, Rotem},
	journal={arXiv preprint arXiv:2410.20363},
	year={2024},
	url={https://arxiv.org/abs/2410.20363}
}

@inproceedings{karaouli2025tsfm,
	title={Are Time Series Foundation Models Ready for Forecasting?},
	author={Karaouli, Nouha and Provoost, Thomas and Verbeke, Wouter},
	booktitle={NeurIPS 2024 Workshop on Recent Advances in Time Series Foundation Models},
	year={2025},
	url={https://arxiv.org/abs/2510.00742}
}

@article{zhang2003time,
	title={Time Series Forecasting Using a Hybrid {ARIMA} and Neural Network Model},
	author={Zhang, G Peter},
	journal={Neurocomputing},
	volume={50},
	pages={159--175},
	year={2003}
}

@book{box1970time,
	title={Time Series Analysis: Forecasting and Control},
	author={Box, George EP and Jenkins, Gwilym M},
	year={1970},
	publisher={Holden-Day},
	address={San Francisco}
}

@article{holt1957forecasting,
	title={Forecasting Seasonals and Trends by Exponentially Weighted Moving Averages},
	author={Holt, Charles C},
	journal={ONR Research Memorandum},
	volume={52},
	year={1957},
	publisher={Carnegie Institute of Technology}
}

@article{winters1960forecasting,
	title={Forecasting Sales by Exponentially Weighted Moving Averages},
	author={Winters, Peter R},
	journal={Management Science},
	volume={6},
	number={3},
	pages={324--342},
	year={1960},
	publisher={INFORMS}
}

@article{tomek1997commodity,
	title={Commodity Futures Prices as Forecasts},
	author={Tomek, William G},
	journal={Review of Agricultural Economics},
	volume={19},
	number={1},
	pages={23--44},
	year={1997},
	publisher={Oxford University Press}
}

@article{sanders2008forecasting,
	title={Forecasting Wheat, Soybean, and Hog Prices: The Contribution of USDA Reports},
	author={Sanders, Dwight R and Manfredo, Mark R},
	journal={Journal of Agricultural and Resource Economics},
	volume={33},
	number={1},
	pages={87--101},
	year={2008}
}

@article{isengildina2011comparison,
	title={A Comparison of USDA and Private Forecasts for Corn, Soybeans, Wheat, and Cotton},
	author={Isengildina-Massa, Olga and MacDonald, Stephen and Xie, Linwood},
	journal={Agribusiness},
	volume={27},
	number={1},
	pages={116--129},
	year={2011}
}

@book{hyndman2018forecasting,
	title={Forecasting: Principles and Practice},
	author={Hyndman, Rob J and Athanasopoulos, George},
	year={2018},
	publisher={OTexts},
	edition={2nd}
}

@article{sims1980macroeconomics,
	title={Macroeconomics and Reality},
	author={Sims, Christopher A},
	journal={Econometrica},
	volume={48},
	number={1},
	pages={1--48},
	year={1980}
}

@article{makridakis2018statistical,
	title={Statistical and Machine Learning Forecasting Methods: Concerns and Ways Forward},
	author={Makridakis, Spyros and Spiliotis, Evangelos and Assimakopoulos, Vassilios},
	journal={PLoS ONE},
	volume={13},
	number={3},
	pages={e0194889},
	year={2018}
}

@article{makridakis2018m4,
	title={The {M4} Competition: Results, Findings, Conclusion and Way Forward},
	author={Makridakis, Spyros and Spiliotis, Evangelos and Assimakopoulos, Vassilios},
	journal={International Journal of Forecasting},
	volume={34},
	number={4},
	pages={802--808},
	year={2018}
}

@article{breiman2001random,
	title={Random Forests},
	author={Breiman, Leo},
	journal={Machine Learning},
	volume={45},
	number={1},
	pages={5--32},
	year={2001}
}

@book{breiman1984cart,
	title={Classification and Regression Trees},
	author={Breiman, Leo and Friedman, Jerome and Stone, Charles J and Olshen, Richard A},
	year={1984},
	publisher={CRC Press}
}

@book{hastie2009elements,
	title={The Elements of Statistical Learning: Data Mining, Inference, and Prediction},
	author={Hastie, Trevor and Tibshirani, Robert and Friedman, Jerome},
	year={2009},
	edition={2nd},
	publisher={Springer},
	address={New York}
}

@inproceedings{chen2016xgboost,
	title={XGBoost: A Scalable Tree Boosting System},
	author={Chen, Tianqi and Guestrin, Carlos},
	booktitle={Proceedings of the 22nd ACM SIGKDD International Conference on Knowledge Discovery and Data Mining},
	pages={785--794},
	year={2016}
}

@article{hochreiter1997long,
	title={Long Short-Term Memory},
	author={Hochreiter, Sepp and Schmidhuber, J{\"u}rgen},
	journal={Neural Computation},
	volume={9},
	number={8},
	pages={1735--1780},
	year={1997}
}

@inproceedings{vaswani2017attention,
	title={Attention is All You Need},
	author={Vaswani, Ashish and Shazeer, Noam and Parmar, Niki and Uszkoreit, Jakob and Jones, Llion and Gomez, Aidan N and Kaiser, {\L}ukasz and Polosukhin, Illia},
	booktitle={Advances in Neural Information Processing Systems (NeurIPS)},
	volume={30},
	pages={5998--6008},
	year={2017}
}

@article{bommasani2021opportunities,
	title={On the Opportunities and Risks of Foundation Models},
	author={Bommasani, Rishi and Hudson, Drew A and Adeli, Ehsan and Altman, Russ and Arora, Simran and von Arx, Sydney and Bernstein, Michael S and Bohg, Jeannette and Bosselut, Antoine and Brunskill, Emma and others},
	journal={arXiv preprint arXiv:2108.07258},
	year={2021},
	url={https://arxiv.org/abs/2108.07258}
}

@article{zulauf2014farm,
	title={Farm Bill Deep Dive: The New ARC and PLC Programs},
	author={Zulauf, Carl and Schnitkey, Gary},
	journal={Farmdoc Daily},
	volume={4},
	number={118},
	year={2014}
}

@article{schnitkey2019plc,
	title={Comparison of PLC and ARC Programs for the 2019 Crop Year},
	author={Schnitkey, Gary and Paulson, Nick and Coppess, Jonathan and Zulauf, Carl},
	journal={Farmdoc Daily},
	volume={9},
	number={156},
	year={2019}
}

@article{diebold1995comparing,
	title={Comparing Predictive Accuracy},
	author={Diebold, Francis X and Mariano, Robert S},
	journal={Journal of Business \& Economic Statistics},
	volume={13},
	number={3},
	pages={253--263},
	year={1995}
}

@article{goodwin2000forecasting,
	title={Forecasting Cattle Prices in the Presence of Structural Change},
	author={Goodwin, Barry K and Schnepf, Randall and Dohlman, Erik},
	journal={Journal of Agricultural and Applied Economics},
	volume={32},
	number={2},
	pages={319--333},
	year={2000}
}

@article{lim2021temporal,
	title={Temporal Fusion Transformers for Interpretable Multi-Horizon Time Series Forecasting},
	author={Lim, Bryan and Arik, Sercan O and Loeff, Nicolas and Pfister, Tomas},
	journal={International Journal of Forecasting},
	volume={37},
	number={4},
	pages={1748--1764},
	year={2021}
}

@inproceedings{oreshkin2020nbeats,
	title={N-BEATS: Neural Basis Expansion Analysis for Interpretable Time Series Forecasting},
	author={Oreshkin, Boris N and Carpov, Dmitri and Chapados, Nicolas and Bengio, Yoshua},
	booktitle={International Conference on Learning Representations (ICLR)},
	year={2020}
}

@article{salinas2020deepar,
	title={DeepAR: Probabilistic Forecasting with Autoregressive Recurrent Networks},
	author={Salinas, David and Flunkert, Valentin and Gasthaus, Jan and Januschowski, Tim},
	journal={International Journal of Forecasting},
	volume={36},
	number={3},
	pages={1181--1191},
	year={2020},
	note={Originally arXiv:1704.04110, 2017}
}

@article{fama1970efficient,
	title={Efficient Capital Markets: A Review of Theory and Empirical Work},
	author={Fama, Eugene F},
	journal={Journal of Finance},
	volume={25},
	number={2},
	pages={383--417},
	year={1970}
}

@article{Working1949theory,
	title={The Theory of Price of Storage},
	author={Working, Holbrook},
	journal={American Economic Review},
	volume={39},
	number={6},
	pages={1254--1262},
	year={1949}
}

@article{wright2011economics,
	title={The Economics of Grain Price Volatility},
	author={Wright, Brian D},
	journal={Applied Economic Perspectives and Policy},
	volume={33},
	number={1},
	pages={32--58},
	year={2011}
}

@book{goodfellow2016deep,
	title={Deep Learning},
	author={Goodfellow, Ian and Bengio, Yoshua and Courville, Aaron},
	year={2016},
	publisher={MIT Press}
}

@article{taylor2018forecasting,
	title={Forecasting at Scale},
	author={Taylor, Sean J and Letham, Benjamin},
	journal={The American Statistician},
	volume={72},
	number={1},
	pages={37--45},
	year={2018}
}

@article{cleveland1990stl,
	title={STL: A Seasonal-Trend Decomposition Procedure Based on Loess},
	author={Cleveland, Robert B and Cleveland, William S and McRae, Jean E and Terpenning, Irma},
	journal={Journal of Official Statistics},
	volume={6},
	number={1},
	pages={3--73},
	year={1990}
}

@inproceedings{devlin2019bert,
	title={{BERT}: Pre-training of Deep Bidirectional Transformers for Language Understanding},
	author={Devlin, Jacob and Chang, Ming-Wei and Lee, Kenton and Toutanova, Kristina},
	booktitle={Proceedings of the 2019 Conference of the North American Chapter of the Association for Computational Linguistics: Human Language Technologies, Volume 1 (Long and Short Papers)},
	pages={4171--4186},
	year={2019}
}

@article{kottapalli2025foundation,
	title={Foundation Models for Time Series: A Survey},
	author={Kottapalli, Siva Rama Krishna and Hubli, Karthik and Chandrashekhara, Sandeep and Jain, Garima and Hubli, Sunayana and Botla, Gayathri and Doddaiah, Ramesh},
	journal={arXiv preprint arXiv:2504.04011},
	year={2025}
}

@article{zhang2024llm,
	title={Large Language Models for Time Series: A Survey},
	author={Zhang, Xiyuan and Chowdhury, Ranak Roy and Gupta, Rajesh K. and Shang, Jingbo},
	journal={arXiv preprint arXiv:2402.01801},
	year={2024}
}

@techreport{hoffman2005forecasting,
	title={Forecasting the Counter-Cyclical Payment Rate for U.S. Corn: An Application of the Futures Price Forecasting Model},
	author={Hoffman, Linwood A},
	institution={USDA Economic Research Service},
	type={Electronic Outlook Report},
	number={FDS-05A-01},
	year={2005},
	month={January}
}

@techreport{hoffman2007forecasting,
	title={Forecasting the Counter-Cyclical Payment Rate for U.S. Corn: An Application of the Futures Price Forecasting Model},
	author={Hoffman, Linwood A and Irwin, Scott H and Toasa, Jesse},
	institution={USDA Economic Research Service},
	type={Outlook Report},
	number={FDS-07D-01},
	year={2007},
	publisher={U.S. Department of Agriculture}
}

@techreport{hoffman2018cotton,
	title={Forecasting the U.S. Season-Average Farm Price of Upland Cotton: Derivation of a Futures Price Forecasting Model},
	author={Hoffman, Linwood A and Meyer, Leslie A},
	institution={USDA Economic Research Service},
	type={Economic Research Report},
	year={2018}
}

@article{poghosyan2025cotton,
	title={Futures-Based Forecasts of Cotton Prices: Beyond Historical Averages},
	author={Poghosyan, Armine and Isengildina-Massa, Olga and Stewart, Shamar L},
	journal={Journal of Agricultural and Applied Economics},
	volume={57},
	pages={114--134},
	year={2025},
	doi={10.1017/aae.2024.36}
}

@article{isengildina2024optimal,
	title={Are USDA Forecasts Optimal? A Systematic Review},
	author={Isengildina-Massa, Olga and Karali, Berna and Irwin, Scott H},
	journal={Journal of Agricultural and Applied Economics},
	volume={56},
	number={3},
	pages={496--527},
	year={2024},
	doi={10.1017/aae.2024.18}
}

@article{etienne2019,
	title={Forecasting Crop Prices Using Futures Prices},
	author={Etienne, Xiaoli L and Irwin, Scott H and Garcia, Philip},
	journal={Agricultural Economics},
	year={2019}
}

@article{figuerola2021,
	title={Oil Price Analysts' Forecasts},
	author={Figuerola-Ferretti, Isabel and Rodr{\'i}guez, Alejandro and Schwartz, Eduardo},
	journal={Journal of Futures Markets},
	volume={41},
	number={9},
	pages={1351--1374},
	year={2021},
	doi={10.1002/fut.22223}
}

@article{hoffman2015,
	title={Forecast Performance of {WASDE} Price Projections for {U.S.} Corn},
	author={Hoffman, Linwood A and Etienne, Xiaoli L and Irwin, Scott H and Colino, Evelyn V and Toasa, Jesse I},
	journal={Agricultural Economics},
	volume={46},
	number={S1},
	pages={157--171},
	year={2015},
	doi={10.1111/agec.12205}
}

@article{kastens1998,
	title={Forecasting Crop Prices: An Evaluation of Futures, Cash, and Composite Forecasts},
	author={Kastens, Terry L and Schroeder, Ted C},
	journal={NCR-134 Conference on Applied Commodity Price Analysis, Forecasting, and Market Risk Management},
	year={1998}
}

@article{manfredo2004,
	title={The Forecasting Performance of Implied Volatility from Live Cattle Options Contracts},
	author={Manfredo, Mark R and Sanders, Dwight R},
	journal={Agribusiness},
	year={2004}
}

@article{roache2011,
	title={Commodity Price Movements in a Decade of Financial Turmoil},
	author={Roache, Shaun K and Reichsfeld, Daniel A},
	journal={IMF Working Paper},
	year={2011}
}

@article{adjemian2020,
	title={Commodity Market Outlook and Price Forecasting},
	author={Adjemian, Michael K and others},
	journal={Agricultural Economics},
	year={2020}
}

@article{colino2010,
	title={Outlook vs. Futures: Three Decades of Evidence in Hog and Cattle Markets},
	author={Colino, Evelyn V and Irwin, Scott H},
	journal={American Journal of Agricultural Economics},
	volume={92},
	number={1},
	pages={1--15},
	year={2010},
	doi={10.1093/ajae/aap018}
}

@article{irwin1994,
	title={The Forecasting Performance of Livestock Futures Prices: A Comparison to USDA Expert Predictions},
	author={Irwin, Scott H and Gerlow, Michael E and Liu, Tsai-Rong},
	journal={Journal of Futures Markets},
	volume={14},
	number={7},
	pages={861--875},
	year={1994}
}

@article{irwin2015,
	title={Long-Term Corn, Soybeans, and Wheat Price Forecasts and the Farm Bill Program Choice},
	author={Irwin, Scott H and Good, Darrel L},
	journal={Farmdoc Daily},
	volume={5},
	number={20},
	year={2015}
}

\clearpage
\appendix

%====================================================================
\section{Data Processing and Validation}
\label{sec:app_data}

%====================================================================
\subsection{Data Sources and Marketing Year Definitions}

The raw data comes from two USDA ERS files downloaded from the Season-Average Price Forecasts database on December 10, 2025. The first file, \texttt{inputdata.csv}, contains 126,849 records of monthly prices received, futures prices, basis values, and marketing percentages; for our analysis, we use only the price received and marketing percentage data spanning 1997 to 2025. The second file, \texttt{outputfc.csv}, contains 11,151 records of MYA price forecasts and actual MYA prices. Our analysis covers 1997 to 2025, corresponding to marketing years 1997 to 2024, with marketing year 2024 being the most recent completed year with final MYA prices available for model evaluation. The data files are available at \url{https://ers.usda.gov/sites/default/files/_laserfiche/DataFiles/53270/inputdata.csv} and \url{https://ers.usda.gov/sites/default/files/_laserfiche/DataFiles/53270/outputfc.csv}.

Marketing years are defined to align with each commodity's harvest and marketing cycle: corn and soybeans (September-August), wheat (June-May), and cotton (August-July).

%====================================================================
\subsection{USDA ERS Forecast Procedure}
\label{sec:app_usda_procedure}

The USDA ERS Season-Average Price Forecasts use a futures-based methodology that combines actual USDA NASS price data with futures-derived forecasts for months where actual prices are unavailable. The procedure: (1) obtains monthly price received data from USDA NASS Agricultural Prices reports; (2) uses daily futures settlement prices of nearby contracts for months without actual data; (3) calculates monthly marketing percentages as 5-year averages for corn, soybeans, and wheat, and 7-year Olympic averages for cotton; (4) calculates monthly basis averages (historical difference between NASS prices and futures prices) using the same averaging periods; (5) constructs monthly price forecasts by adding basis average to futures settlement price; (6) creates composite monthly prices using actual NASS data where available and forecasted prices otherwise; (7) calculates monthly MYA weights as the product of composite prices and marketing percentages; (8) sums monthly weights to obtain the final MYA forecast:
\begin{align*}
   \text{MYA} &= \sum_{m=1}^{12} \text{Composite Price}_m \times \text{Marketing Percentage}_m.
\end{align*}

%====================================================================
\section{Time Series Foundation Model Details}
\label{sec:app_tsfm}

Table~\ref{tab:tsfm_summary} summarizes the five time series foundation models evaluated in this study.

\begin{table}[ht!]
   \centering
   \caption{Summary of Time Series Foundation Models}
   \label{tab:tsfm_summary}
   \footnotesize
   \begin{tabular}{llllll}
      \toprule
      \textbf{Model} & \textbf{Provider} & \textbf{Params} & \textbf{Architecture} & \textbf{Model Identifier} \\
      \midrule
      Time-MoE & Maple728 & 113M & MoE Transformer & \texttt{Maple728/TimeMoE-50M} \\
      Chronos & Amazon & 201M & Language Model & \texttt{amazon/chronos-t5-base} \\
      Chronos-2 & Amazon & 119M & T5 Encoder & \texttt{amazon/chronos-2} \\
      TimesFM 2.5 & Google & 231M & Decoder-only & \texttt{google/timesfm-2.5-200m-pytorch} \\
      Moirai-2 & Salesforce & 11M & Decoder-only & \texttt{Salesforce/moirai-2.0-R-small} \\
      \bottomrule
   \end{tabular}
   \vskip 2mm
   \begin{minipage}{0.98\textwidth}
      {\footnotesize \textit{Notes:} Summary of foundation models evaluated in zero-shot mode. Params = total number of parameters in the released checkpoint, counted directly from the downloaded model weights. MoE = Mixture-of-Experts. For Time-MoE, the 113M total reflects 8 experts per feed-forward layer of which only 2 are activated per token, so roughly 33M parameters are active in any single forward pass; the ``50M'' in the checkpoint name refers to this sparsely activated scale. All models are accessed through their PyTorch implementations via HuggingFace, except Chronos-2, whose weights are loaded from the AutoGluon S3 mirror.\par}
   \end{minipage}
\end{table}

Table~\ref{tab:tsfm_config} details the inference configuration used for each foundation model in our evaluation.

\begin{table}[ht!]
   \centering
   \caption{TSFM Inference Configuration}
   \label{tab:tsfm_config}
   \footnotesize
   \begin{tabular}{lllll}
      \toprule
      \textbf{Model} & \textbf{Context Length} & \textbf{Max Horizon} & \textbf{Normalization} & \textbf{Output Type} \\
      \midrule
      Time-MoE & Max 4,096 & Any (AR) & External (z-score) & Point (mean) \\
      Chronos & Max 512 & 64 & Built-in & Median of 20 samples \\
      Chronos-2 & Max 8,192 & 1,024 & Built-in & Quantiles (0.1, 0.5, 0.9) \\
      TimesFM 2.5 & Max 16,384 & Any (AR) & Built-in & Point forecast \\
      Moirai-2 & Max 4,096 & Any (AR) & Built-in & Quantiles (9 levels) \\
      \bottomrule
   \end{tabular}
   \vskip 2mm
   \begin{minipage}{0.98\textwidth}
      {\footnotesize \textit{Notes:} Configuration details for zero-shot inference. Context length indicates maximum historical observations the model can process. Max horizon shows model capacity; ``Any (AR)'' indicates autoregressive generation that supports arbitrary forecast horizons by iteratively generating predictions. We use 12-month horizon for all models. Time-MoE requires external z-score normalization before inference; all other models handle normalization internally. For probabilistic models, we use the median (0.5 quantile) as the point forecast.\par}
   \end{minipage}
\end{table}

%====================================================================
\section{Cross-Validation Design and Hyperparameter Optimization}
\label{sec:app_cv}

%====================================================================
\subsection{Evaluation 1: Monthly Price Forecasting}

The objective is to evaluate model accuracy at predicting individual monthly prices throughout the marketing year. The procedure is as follows:
\begin{enumerate}
   \item For each split $s$ and commodity $c$:
   \begin{itemize}
      \item Input: Historical monthly prices up to start of test year
      \item Output: 12 monthly price forecasts for test year
      \item Horizon: Fixed 12-month ahead forecast
   \end{itemize}

\item Calculate monthly-level metrics:
\begin{align}
   \text{MAE}_{\text{monthly}} &= \frac{1}{12} \sum_{t=1}^{12} |P_t^{\text{actual}} - P_t^{\text{forecast}}| \\
   \text{RMSE}_{\text{monthly}} &= \sqrt{\frac{1}{12} \sum_{t=1}^{12} (P_t^{\text{actual}} - P_t^{\text{forecast}})^2} \\
   \text{MAPE}_{\text{monthly}} &= \frac{100}{12} \sum_{t=1}^{12} \frac{|P_t^{\text{actual}} - P_t^{\text{forecast}}|}{P_t^{\text{actual}}}
\end{align}

\item Aggregate across all splits and commodities:
\begin{align}
   \text{Overall MAE} &= \frac{1}{N} \sum_{i=1}^{N} \text{MAE}_i,
\end{align}
where $N = 64$ (total number of split-commodity combinations)
\end{enumerate}

Monthly MAE measures average forecast error across individual months. This metric captures the model's ability to track month-to-month price movements and seasonal patterns.

%====================================================================
\subsection{Evaluation 2: Marketing Year Average (MYA) Forecasting and USDA Comparison}

The objective is to evaluate model accuracy at predicting the policy-relevant Marketing Year Average price and compare performance against USDA ERS operational forecasts. The procedure is as follows:
\begin{enumerate}
   \item For each split $s$ and commodity $c$:
   \begin{itemize}
      \item Generate 12 monthly forecasts (same as Evaluation 1)
      \item Aggregate to MYA using marketing percentages as weights:
      \begin{align}
         \text{MYA}^{\text{forecast}} &= \sum_{t=1}^{12} P_t^{\text{forecast}} \times w_t,
      \end{align}
      where $w_t$ are the marketing percentages defined in Section~\ref{sec:mya_prices}
   \end{itemize}

\item Calculate MYA-level metrics:
\begin{align}
   \text{MAE}_{\text{MYA}} &= |\text{MYA}^{\text{actual}} - \text{MYA}^{\text{forecast}}| \\
   \text{RMSE}_{\text{MYA}} &= \sqrt{(\text{MYA}^{\text{actual}} - \text{MYA}^{\text{forecast}})^2} \\
   \text{MAPE}_{\text{MYA}} &= 100 \times \frac{|\text{MYA}^{\text{actual}} - \text{MYA}^{\text{forecast}}|}{\text{MYA}^{\text{actual}}}
\end{align}

\item Aggregate using two-step commodity-based averaging:
\begin{align}
   \text{Commodity MAE}_c &= \frac{1}{N_c} \sum_{i \in c} \text{MAE}_{\text{MYA},i}, \quad \text{for each commodity } c \\
   \text{Overall MYA MAE} &= \frac{1}{4} \sum_{c=1}^{4} \text{Commodity MAE}_c.
\end{align}
This two-step process first averages MAE across all matched years within each commodity, then averages these four commodity-specific MAEs. This ensures equal weight for each commodity regardless of the number of available years.
\end{enumerate}

MYA aggregation causes error cancellation, as overforecasts in some months offset underforecasts in others. Consequently, $\text{MAE}_{\text{MYA}} < \text{MAE}_{\text{monthly}}$ systematically. This is why we must compare USDA's MYA forecasts to our MYA forecasts (not monthly forecasts) for apples-to-apples comparison.

USDA forecasts MYA directly using the futures-basis approach, while our models forecast 12 monthly prices and then aggregate to MYA. Both are evaluated on the same test years (2009-2024 for corn, soybeans, and wheat; 2019-2024 for cotton) using MYA MAE, RMSE, and MAPE metrics. For cotton, USDA forecasts are only available from 2019 onwards, so all models are evaluated on the 2019-2024 period only (6 years, splits 11-16) for fair comparison. Other commodities use the full 2009-2024 period (16 years, splits 1-16).

%====================================================================
\subsection{Cross-Validation Split Details}

Our 16 expanding window splits are structured as follows. Each split uses an expanding training window (growing by 1 year per split) with fixed 2-year validation and 1-year test windows:

\begin{table}[ht!]
   \centering
   \caption{Complete Cross-Validation Split Structure with Record Counts}
   \label{tab:cv_splits}
   \footnotesize
   \begin{tabular}{ccccccc}
   \toprule
   Split & Train Period & Train N & Val Period & Val N & Test Period & Test N \\
   \midrule
   1 & 1997-2006 & 480 & 2007-2008 & 96 & 2009 & 48 \\
   2 & 1997-2007 & 528 & 2008-2009 & 96 & 2010 & 48 \\
   3 & 1997-2008 & 576 & 2009-2010 & 96 & 2011 & 48 \\
   4 & 1997-2009 & 624 & 2010-2011 & 96 & 2012 & 48 \\
   5 & 1997-2010 & 672 & 2011-2012 & 96 & 2013 & 48 \\
   6 & 1997-2011 & 720 & 2012-2013 & 96 & 2014 & 48 \\
   7 & 1997-2012 & 768 & 2013-2014 & 96 & 2015 & 48 \\
   8 & 1997-2013 & 816 & 2014-2015 & 96 & 2016 & 48 \\
   9 & 1997-2014 & 864 & 2015-2016 & 96 & 2017 & 48 \\
   10 & 1997-2015 & 912 & 2016-2017 & 96 & 2018 & 48 \\
   11 & 1997-2016 & 960 & 2017-2018 & 96 & 2019 & 48 \\
   12 & 1997-2017 & 1,008 & 2018-2019 & 96 & 2020 & 48 \\
   13 & 1997-2018 & 1,056 & 2019-2020 & 96 & 2021 & 48 \\
   14 & 1997-2019 & 1,104 & 2020-2021 & 96 & 2022 & 48 \\
   15 & 1997-2020 & 1,152 & 2021-2022 & 96 & 2023 & 48 \\
   16 & 1997-2021 & 1,200 & 2022-2023 & 96 & 2024 & 48 \\
   \bottomrule
   \end{tabular}
   \vskip 2mm
   \begin{minipage}{0.98\textwidth}
      {\footnotesize \textit{Notes}: All periods refer to marketing years, which span calendar years (e.g., marketing year 2024 runs from September 2024 to August 2025 for corn and soybeans). Training set grows from 480 observations (10 years) to 1,200 observations (25 years), ensuring robust evaluation across different time periods while maintaining temporal ordering. Each year contributes 48 observations (12 months $\times$ 4 commodities).\par}
   \end{minipage}
\end{table}

The split design has several key features. The expanding window means training data accumulates over time, mimicking real forecasting scenarios. The minimum training size of 10 years (480 observations) provides sufficient history for seasonal patterns. Fixed validation and test windows ensure that 2-year validation and 1-year test periods remain constant. Temporal ordering strictly respects time series structure to prevent data leakage. The design evaluates performance across 16 distinct years (2009-2024).

%====================================================================
\subsection{Complete Hyperparameter Specifications}

Table \ref{tab:hyperparameters} provides complete hyperparameter specifications for all 17 models evaluated in this study. For models with grid search, we report the search space and selection method. For foundation models, we report the inference settings used.

\begin{table}[ht!]
   \centering
   \caption{Complete Hyperparameter Specifications for All Models}
   \label{tab:hyperparameters}
   \scriptsize
   \begin{tabular}{p{2.2cm}p{3.5cm}p{7cm}p{2cm}}
   \toprule
   \textbf{Model} & \textbf{Key Parameters} & \textbf{Grid Search Space / Settings} & \textbf{Selection} \\
   \midrule
   \multicolumn{4}{l}{\textit{\textbf{Traditional Time Series Models}}} \\
   \midrule
   Naive & None & No hyperparameters & N/A \\
   \midrule
   Seasonal Naive & seasonal\_period & 12 (fixed) & N/A \\
   \midrule
   SARIMA & p, d, q, P, D, Q, s & p,q,P,Q $\in \{0,1,2\}$; d,D $\in \{0,1\}$; s=12 & AIC \\
   \midrule
   Exp Smoothing & trend, seasonal & trend $\in$ \{add, mul, none\}; seasonal $\in$ \{add, mul, none\} (9 combos) & Val RMSE \\
   \midrule
   STL & seasonal\_window, trend\_window & seasonal $\in \{7,13,25,35\}$; trend $\in \{$None$,13,25,51\}$ (16 combos) & Val RMSE \\
   \midrule
   Prophet & changepoint\_prior, seasonality\_prior, mode, cp\_range & cp\_prior $\in \{0.001,0.01,0.05,0.1,0.5,1.0\}$; seas\_prior $\in \{0.01,0.1,1.0,10.0,20.0\}$; mode $\in$ \{add,mul\}; cp\_range $\in \{0.8,0.9,0.95\}$ & Val RMSE \\
   \midrule
   \multicolumn{4}{l}{\textit{\textbf{Machine Learning Models}}} \\
   \midrule
   Random Forest & n\_lags, n\_estimators, max\_depth & n\_lags $\in \{6,12,18\}$; n\_estimators $\in \{100,200\}$; max\_depth $\in \{10,15,20\}$ & Val RMSE \\
   \midrule
   XGBoost & n\_lags, n\_estimators, learning\_rate & n\_lags $\in \{6,12,18\}$; n\_estimators $\in \{100,200\}$; lr $\in \{0.05,0.1\}$; max\_depth=6; subsample=0.8; colsample=0.8 & Val RMSE \\
   \midrule
   \multicolumn{4}{l}{\textit{\textbf{Deep Learning Models}}} \\
   \midrule
   LSTM & seq\_length, hidden\_size, num\_layers, epochs & seq\_len $\in \{12,24\}$; hidden $\in \{32,64\}$; layers=1; batch=16; lr=0.001; epochs $\in \{30,50\}$ & Val RMSE \\
   \midrule
   N-BEATS & n\_blocks, mlp\_units, max\_steps & n\_blocks=[1,1]; mlp\_units=[[64,64]]; harmonics=1; polynomials=2; max\_steps=30 (simplified) & Fixed \\
   \midrule
   TFT & hidden\_size, n\_head, max\_steps & hidden\_size=32; n\_head=2; max\_steps=50 (simplified) & Fixed \\
   \midrule
   DeepAR & lstm\_hidden, lstm\_layers, max\_steps & hidden $\in \{32,64\}$; layers $\in \{1,2\}$; max\_steps $\in \{30,50\}$ & Val RMSE \\
   \midrule
   \multicolumn{4}{l}{\textit{\textbf{Foundation Models (Zero-Shot)}}} \\
   \midrule
   Chronos & N/A & N/A & Fixed \\
   \midrule
   Chronos-2 & N/A & N/A & Fixed \\
   \midrule
   TimesFM 2.5 & N/A & N/A & Fixed \\
   \midrule
   Time-MoE & N/A & N/A & Fixed \\
   \midrule
   Moirai-2 & N/A & N/A & Fixed \\
   \bottomrule
   \end{tabular}
   \vskip 2mm
   \begin{minipage}{0.98\textwidth}
      {\footnotesize \textit{Notes}: Complete hyperparameter specifications for all 17 models. Traditional and ML models with multiple configurations select best parameters via validation RMSE. Foundation models use zero-shot inference with pre-trained weights and fixed settings (no grid search). Deep learning models use Adam optimizer with early stopping. "Simplified" architectures for N-BEATS and TFT were necessary to prevent overfitting on limited agricultural data (200-250 training samples). Random seeds set to 42 for reproducibility.\par}
   \end{minipage}
\end{table}

%====================================================================
\section{Data Leakage Evaluation Framework}
\label{sec:app_simulation}

A critical concern in evaluating Time Series Foundation Models (TSFMs) is distinguishing between genuine pattern learning and memorization of training data. As \citet{meyer2025tsfm} highlight, the field faces significant challenges from "risks of information leakage due to overlapping and obscure datasets" and "memorization of global patterns," yet lacks established evaluation methodologies to address these concerns. This appendix presents a comprehensive framework for evaluating data leakage risks through multiple complementary approaches.

The fundamental challenge lies in the opacity of TSFM training data. While foundation models demonstrate impressive zero-shot performance, their training corpora often include vast collections of time series data that may overlap with evaluation benchmarks. Agricultural price data, being publicly available and economically significant, represents a particularly high-risk domain for potential contamination. We address this concern through three methodological approaches: training data documentation, temporal stratification, and synthetic benchmark controls.

%====================================================================
\subsection{Training Data Documentation}

Understanding the potential for data contamination requires systematic documentation of known pretraining corpora for each TSFM. Our investigation reveals varying levels of transparency across foundation models:

Chronos (published March 2024) was trained on a diverse collection including the Monash Time Series Forecasting Archive, which contains over 30,000 time series from various domains. While USDA ERS price data are not explicitly listed in their training documentation, the Monash Archive includes agricultural and commodity datasets that may exhibit similar statistical properties to our evaluation data. Based on the March 2024 publication date, training likely concluded by mid-2023.

TimesFM 2.5 (submitted October 2023, final version April 2024) utilized Google's internal time series collections, with limited public documentation of specific datasets. The model was trained on "real-world time series data" including financial, retail, and web traffic data. Agricultural commodity prices, being economically significant and publicly available, could plausibly exist in such collections. Training data cutoff is estimated at mid-2023 based on submission timeline.

Time-MoE (submitted September 2024, accepted ICLR 2025) employed a mixture-of-experts architecture trained on their newly introduced Time-300B dataset spanning over 9 domains and encompassing over 300 billion time points. Their documentation mentions "diverse real-world datasets" but provides limited specifics about agricultural or commodity data inclusion. Training appears to have concluded by mid-2024 based on submission date.

Chronos-2 (submitted October 2024) represents an extension of the original Chronos with expanded multivariate capabilities. The model builds upon the original Chronos training corpus with additional datasets, suggesting training completion by late 2024. Moirai-2 (submitted November 2024) represents the most recent model in our evaluation, with training likely extending into late 2024 based on submission timeline.

None of the models explicitly list USDA ERS price data in their documented training sets. However, agricultural price data's widespread availability and inclusion in major repositories (Monash Archive, FRED, World Bank) suggests potential indirect exposure. We address this concern through simulation analysis using synthetically generated price series.

%====================================================================
\subsection{Simulation Analysis}

We generate synthetic agricultural price series that preserve real data's statistical properties while ensuring zero contamination risk. If TSFMs have learned genuine agricultural price dynamics, their performance on synthetic data should be comparable to real data performance. Dramatically different performance would suggest memorization of specific historical sequences rather than learning of underlying economic patterns. All models use identical configurations as in the main evaluation (see Table \ref{tab:hyperparameters}), ensuring that performance differences reflect data characteristics rather than model specification changes.

%====================================================================
\subsubsection{Synthetic Data Generation}

We employ Gaussian Process (GP) regression to generate 100 synthetic price series per commodity (400 total) matching USDA data's statistical properties. The GP framework uses three kernel components:

\textbf{Periodic Kernel (Seasonality):} Agricultural prices exhibit strong seasonal patterns due to planting, growing, and harvest cycles. We model this using a periodic kernel:
\begin{align}
   k_{\text{per}}(t, t') &= \sigma_p^2 \exp\left(-\frac{2\sin^2(\pi|t-t'|/12)}{\ell_p^2}\right),
\end{align}
where $\sigma_p^2$ controls seasonal amplitude and $\ell_p$ determines seasonal smoothness. Parameters are calibrated to match the seasonal variance observed in USDA corn, soybean, and wheat prices.

\textbf{RBF Kernel (Long-term Trends):} Long-term price trends reflect macroeconomic factors, technological progress, and structural market changes. We capture these using a radial basis function (RBF) kernel:
\begin{align}
   k_{\text{rbf}}(t, t') &= \sigma_r^2 \exp\left(-\frac{|t-t'|^2}{2\ell_r^2}\right),
\end{align}
where $\sigma_r^2$ controls trend magnitude and $\ell_r$ determines trend persistence. Parameters are fitted to match the long-term volatility characteristics of agricultural commodities.

\textbf{Noise Kernel (Short-term Volatility):} Agricultural prices exhibit significant short-term volatility due to weather events, policy announcements, and market speculation. We model this using a white noise kernel:
\begin{align}
   k_{\text{noise}}(t, t') &= \sigma_n^2 \delta_{t,t'},
\end{align}
where $\sigma_n^2$ matches the residual variance after removing seasonal and trend components from real USDA data.

\textbf{Combined Kernel:} The full covariance function combines all components:
\begin{align}
   k(t, t') &= k_{\text{per}}(t, t') + k_{\text{rbf}}(t, t') + k_{\text{noise}}(t, t').
\end{align}

%====================================================================
\subsubsection{Parameter Calibration}

We calibrate GP parameters to match key statistical properties of USDA agricultural price series. Periodic kernel parameters are fitted to capture the 15-25\% seasonal price variation typical of agricultural commodities, reflecting the natural cycles of planting, growing, and harvest seasons. The RBF length scale is calibrated to match the 2-5 year trend cycles observed in commodity markets, capturing macroeconomic influences and structural market changes. Noise variance is set to reproduce the 20-40\% annual volatility characteristic of agricultural prices, accounting for weather events, policy announcements, and market speculation. Finally, the combined kernel structure is designed to match the 0.85-0.95 first-order autocorrelation typical of monthly price series, ensuring realistic temporal dependencies in the synthetic data.

%====================================================================
\subsubsection{Simulation Results}

We generate 100 synthetic price series per commodity (400 total) matching the statistical properties of USDA data. Table \ref{tab:synthetic_results} presents comprehensive evaluation results across all 17 models on 400 synthetic series (100 series $\times$ 4 commodities).

\begin{table}[ht!]
   \centering
   \caption{Synthetic Benchmark Results: Model Performance Rankings}
   \label{tab:synthetic_results}
   \footnotesize
   \begin{tabular}{llcccccc}
      \toprule
      Rank & Model & Category & RMSE & MAE & MAPE & SMAPE & Time (s) \\
      \midrule
      1 & Chronos-2 & TSFM & 0.339 & 0.278 & 6.19 & 6.20 & 0.23 \\
      2 & Moirai-2 & TSFM & 0.342 & 0.281 & 6.33 & 6.43 & 0.44 \\
      3 & TimesFM 2.5 & TSFM & 0.344 & 0.282 & 6.32 & 6.41 & 0.85 \\
      4 & Exp. Smoothing & Traditional & 0.344 & 0.283 & 6.66 & 6.97 & 0.07 \\
      5 & Chronos & TSFM & 0.375 & 0.311 & 6.82 & 6.82 & 1.51 \\
      \midrule
      6 & SARIMA & Traditional & 0.393 & 0.326 & 7.68 & 8.20 & 0.40 \\
      7 & Time-MoE & TSFM & 0.402 & 0.337 & 7.45 & 7.25 & 1.94 \\
      8 & STL & Traditional & 0.402 & 0.333 & 7.82 & 8.15 & 0.22 \\
      9 & Prophet & Traditional & 0.423 & 0.357 & 8.44 & 8.98 & 0.08 \\
      10 & LSTM & Deep Learning & 0.427 & 0.360 & 7.56 & 7.54 & 6.11 \\
      \midrule
      11 & Naive & Traditional & 0.429 & 0.364 & 7.85 & 7.80 & 0.00 \\
      12 & Random Forest & ML & 0.434 & 0.366 & 7.66 & 7.62 & 2.96 \\
      13 & N-BEATS & Deep Learning & 0.435 & 0.366 & 8.52 & 8.79 & 0.28 \\
      14 & TFT & Deep Learning & 0.439 & 0.374 & 9.12 & 8.46 & 5.22 \\
      15 & XGBoost & ML & 0.459 & 0.388 & 8.07 & 8.05 & 3.55 \\
      16 & Seasonal Naive & Traditional & 0.490 & 0.409 & 8.73 & 8.64 & 0.00 \\
      17 & DeepAR & Deep Learning & 1.003 & 0.954 & 17.65 & 16.74 & 10.94 \\
      \bottomrule
   \end{tabular}
   \vskip 2mm
   \begin{minipage}{0.98\textwidth}
      {\footnotesize \textit{Notes:} Results averaged across 400 synthetic series (100 per commodity). RMSE and MAE in normalized units. MAPE and SMAPE in percentages. Inference time in seconds per series. Models ranked by RMSE.\par}
   \end{minipage}
\end{table}

\textbf{Key Findings.} TSFMs dominate the top rankings, with Chronos-2, Moirai-2, and TimesFM 2.5 occupying the top three positions, demonstrating that foundation models have learned generalizable time series patterns rather than memorizing specific historical sequences. The performance gap between TSFMs and traditional methods is modest but consistent -- Chronos-2 achieves 1.5\% lower RMSE than Exponential Smoothing (0.339 vs. 0.344). Deep learning models trained from scratch perform poorly, with DeepAR ranking last, confirming that limited training data (144 months per series) is insufficient for neural architectures while pre-trained foundation models leverage knowledge from massive training corpora.

%====================================================================
\section{Pairwise Diebold-Mariano Test Results}
\label{sec:app_dm_tests}

Table \ref{tab:dm_pairwise} presents the complete pairwise Diebold-Mariano test results for all TSFM versus baseline model comparisons in monthly price forecasting evaluation. Each cell shows the direction of the comparison (+ indicates TSFM outperforms the corresponding model) and statistical significance level.

\begin{table}[ht!]
   \centering
   \caption{Pairwise Diebold-Mariano Test Results: TSFMs vs Baseline Models}
   \label{tab:dm_pairwise}
   \resizebox{\textwidth}{!}{%
   \begin{tabular}{lcccccccccccc}
      \toprule
      & Naive & S.Naive & SARIMA & ETS & STL & LSTM & N-BEATS & TFT & DeepAR & RF & XGB & Prophet \\
      \midrule
      Chronos & $-$ & +** & + & +* & +*** & +** & +** & +*** & +*** & + & + & +*** \\
      Chronos-2 & +* & +*** & +** & +** & +*** & +*** & +*** & +*** & +*** & +*** & +** & +*** \\
      TimesFM 2.5 & + & +*** & +** & +** & +*** & +*** & +*** & +*** & +*** & +*** & +** & +*** \\
      Time-MoE & +* & +*** & +** & +** & +*** & +*** & +*** & +*** & +*** & +*** & +*** & +*** \\
      Moirai-2 & + & +*** & +* & +* & +*** & +*** & +*** & +*** & +*** & +*** & +** & +*** \\
      \bottomrule
   \end{tabular}
   }
   \vskip 2mm
   \begin{minipage}{0.98\textwidth}
      {\footnotesize \textit{Notes:} Diebold-Mariano tests using absolute percentage errors pooled across all commodities and forecast horizons. + indicates TSFM outperforms baseline (lower APE); $-$ indicates baseline outperforms TSFM. Significance levels: * $p<0.10$, ** $p<0.05$, *** $p<0.01$. S.Naive = Seasonal Naive, ETS = Exponential Smoothing, RF = Random Forest, and XGB = XGBoost.\par}
   \end{minipage}
\end{table}

The pairwise results reveal systematic patterns. All five TSFMs significantly outperform deep learning models trained from scratch (LSTM, N-BEATS, TFT, DeepAR) at the 1\% level, reflecting the fundamental advantage of pre-trained representations over limited-data training. Against traditional methods, TSFMs show consistent advantages: all five significantly outperform STL and Seasonal Naive at the 1\% level, while comparisons against Naive, SARIMA, and Exponential Smoothing show mixed significance. The single negative result (Chronos vs Naive) reflects Chronos's relatively weaker performance among TSFMs. Against machine learning models, TSFMs demonstrate strong dominance, with most comparisons significant at the 1\% or 5\% level.

\end{document}